\documentclass[%
twocolumn,amsfonts,showpacs,superscriptaddress,
 amsmath,amssymb,
]{revtex4-1}

\usepackage{dutchcal}
\usepackage{graphicx}
\usepackage{bm}
\usepackage{physics}
\usepackage{color}

\newcommand{\bc}{\begin{center}}
\newcommand{\ec}{\end{center}}
\def\ba#1{\begin{array}{#1}\displaystyle}
\newcommand{\ea}{\end{array}}

\newcommand{\beq}{\begin{equation}}
\newcommand{\eeq}{\end{equation}}
\newcommand{\beqa}{\begin{eqnarray}}
\newcommand{\eeqa}{\end{eqnarray}}

\newcommand{\bi}{\begin{itemize}}
\newcommand{\ei}{\end{itemize}}

\renewcommand{\vec}[1]{\boldsymbol{#1}}
\newcommand{\exv}[1]{\expectationvalue{#1}}

\def\b#1{\bar{#1}}
\def\frc#1#2{\frac{#1}{#2}}

\newcommand{\p}{\partial}

\newcommand{\ri}{{\rm i}}

\newcommand{\Or}{{\cal O}}

\newcommand{\varep}{\varepsilon}

\begin{document}


\title{Non-Equilibrium Dynamics and Weakly Broken Integrability}
\author{Joseph Durnin}
\affiliation
{Department of Mathematics, King's College London, Strand WC2R 2LS, UK}
\author{M. J. Bhaseen}
\affiliation
{Department of Physics, King's College London, Strand WC2R 2LS, UK}
\author{Benjamin Doyon}
\affiliation
{Department of Mathematics, King's College London, Strand WC2R 2LS, UK}
%
%
\begin{abstract}
  Motivated by dynamical experiments on cold atomic gases, we develop
  a quantum kinetic approach to weakly perturbed integrable models out
  of equilibrium. Using the exact matrix elements of the underlying
  integrable model we establish an analytical approach to real-time
  dynamics. The method addresses a broad range of timescales, from the
  intermediate regime of pre-thermalization to late-time
  thermalization. Predictions are given for the time-evolution of
  physical quantities, including effective temperatures and
  thermalization rates. The approach provides conceptual links between
  perturbed quantum many-body dynamics and classical
  Kolmogorov-Arnold-Moser (KAM) theory. In particular, we identify a
  family of perturbations which do not cause thermalization in the weakly perturbed regime.
\end{abstract}

\maketitle

Conservation laws play a ubiquitous role in constraining the dynamics
of complex many-body systems. This is especially true in
low-dimensional integrable systems, where their proliferation gives
rise to rich phenomena. A striking example is provided by the quantum
Newton's cradle experiment \cite{kinoshita}, which shows the absence
of thermalization over long timescales. The impact of conservation
laws in this so-called pre-thermalization regime is directly encoded
via a Generalized Gibbs Ensemble (GGE)
\cite{PhysRevLett.98.050405,rigol_breakdown_2009,cassidy_generalized_2011,gring_relaxation_2012,PhysRevLett.115.157201,Langen207}:
each conserved quantity is associated with its own effective
temperature, leading to anomalous thermalization. This has stimulated
a wealth of theoretical activity, including the recent extension of
hydrodynamics
\cite{hydro1,hydro2,hydro3,hydro4,BDLS} to integrable systems
\cite{PhysRevX.6.041065,PhysRevLett.117.207201,PhysRevB.97.045407} and
its application to experiment
\cite{cauxnewton,Schemmerghd}.
For recent reviews exploring the exotic dynamics of isolated quantum integrable systems see \cite{eisertreview,1742-5468-2016-6-064002,VassMoo16,IlievskietalQuasilocal,1742-5468-2016-6-064005,DallesioETHreview2016,GogolinEquilibration2016,DoyonLecture2020}.

Despite recent advances in the understanding of integrable systems,
real physical systems always contain perturbations. These may
influence and destabilize the integrable dynamics, but their effect is
hard to quantify. In the classical domain, the effect of weak
perturbations is encoded in KAM theory
\cite{lazutkin_kam_1993}, which describes the persistence of
quasi-periodic orbits under small perturbations. In the quantum
many-body domain, the scenario of pre-thermalization followed by slow thermalization has been widely studied in this context
\cite{mazets_breakdown_2008,moeckel_interaction_2008,marcuzzi_prethermalization_2013,essler_quench_2014,lux_hydrodynamic_2014,mierzejewski_approximate_2015,bertini_prethermalization_2015,nessi_glass-like_2015,bertini_thermalization_2016,biebl_thermalization_2017,albaPrethermalization2017,lange_time-dependent_2018,van_regemortel_prethermalization_2018,mallayya_prethermalization_2018,tang_thermalization_2018,vas19};
for recent reviews see \cite{Mori_thermalization_2018,mitra_quantum_2018}. However,
insights analagous to KAM theory have been hard to establish, and
many experimentally and conceptually relevant questions
remain. To what extent does quantum integrability survive in the
presence of weak perturbations? How can we quantify and organize the
dynamical effects of integrability destroying interactions? What are
the relevant timescales?

In this paper we address these questions by developing a quantum
kinetic approach to weakly perturbed integrable models out of
equilibrium.
We show that the dynamics of physical observables from
short to long timescales can be described using the exact matrix
elements of the underlying integrable model. Our findings are
illustrated by numerical evaluation of the key formulae, including the
time-evolution of the average densities, quasiparticle distributions,
and effective temperatures. Embedding the kinetic approach into a general
theory, we identify dynamical response functions which
encode the timescales of thermalization. We also find a family of integrability-breaking, KAM-like perturbations, which do not lead to thermalization in the weakly coupled regime.
A notable insight which emerges from our analysis is that, in one
spatial dimension, thermalization and hydrodynamic diffusion are
controlled by distinct families of processes, which we
characterize. Our findings also provide the integrability destroying
corrections to the Euler hydrodynamics of integrable systems.

{\em Setup}.--- We consider the general scenario in which a
spatially homogeneous one-dimensional integrable system, described by
Hamiltonian $ H_0$, is perturbed by an extensive integrability
destroying term $ V=\int \dd x\,v(x)$. The resulting Hamiltonian
is given by $ H= H_0+\lambda  V$, where $\lambda$ controls
the strength of the perturbation. The Hamiltonian $ H_0$ is
characterized by an infinite number of mutually commuting conserved
quantities $ Q_i$, $i=0,1,2,\dots$, including the momentum
$ P= Q_1$ and the Hamiltonian $ H_0= Q_2$. In the
perturbed system only two conserved quantities remain: the
total energy $ H$ and the total momentum $ P$.

In order to explore the dynamics of the non-integrable Hamiltonian $
H$ we consider a quantum quench from an initial state which is
stationary under $H_0$, but which evolves under the dynamics of
$H$. In light of the integrability of $ H_0$ it is natural to take a
GGE as the initial state, whose density matrix is given by $ \rho_0=
Z^{-1} e^{-\sum_{i}\beta_i { Q}_i}$. Here $Z={\rm Tr}(e^{-\sum_{i}\beta_i { Q}_i})$ and the
$\beta_i$ are the inverse effective temperatures associated with each
conserved quantity $Q_i$. These are the most general states that
maximize entropy with respect to all of the extensive conserved
quantities of $H_0$; they therefore provide natural initial states for
studying the dynamics of perturbed integrable systems.

The quench setup described above is well-suited to studying
thermalization. At long times, it is expected that expectation values of local
observables $\expectationvalue{\Or (x,t)} = \Tr\big[\rho_0 e^{\ri H
    t}\Or(x)e^{-\ri H t}\big]$ tend to the value they
would take in a boosted thermal ensemble described by $H$ and
$P$. Explicitly, $\lim_{t\to\infty} \expectationvalue{\Or(x,t)} =
Z_s^{-1}\Tr\big[e^{-\beta_{\rm s} (H -\nu_{\rm s} P)}\Or(x)\big]$,
where the stationary values $\beta_{\rm s}$ and $\nu_{\rm s}$ are
uniquely fixed by $\expectationvalue{ H}$ and $\expectationvalue{
  P}$. Thermalization is proven rigorously in various situations
\cite{rieraThermalization2012,sirkerLocality2014,muellerThermalization2015,Doyon2017},
and if it occurs it does so for any perturbation strength
$\lambda$. From a physical perspective however, the most important
questions are to what extent integrability still plays a role at
finite times, and how the system reaches thermalization. For small
perturbations, it could be expected that integrability strongly
influences these processes, and constrains the dominant physics.

{\em Dynamics of Charges.}--- To see the effects of the integrability-breaking term, it is instructive to examine the time-evolution of the charges $ Q_i$ under the 
 Hamiltonian $H$. To lowest order in
$\lambda$, the time-evolution of the corresponding charge densities
$ q_i(x,t)$, can be computed within second order perturbation theory:
\beq \label{eqn_charge_general} \partial_t\expectationvalue{ q_i
  (0,t)}= \lambda^2 \int_{-t}^t \dd s \expectationvalue{[V^0(s),Q_i]
  v(0)}^{\rm c},\eeq where
here and throughout we set $\hbar=k_B=1$, and we denote the connected correlation function by $\langle\dots\rangle^{\rm c}$. Time-evolution on the left-hand
side is with respect to the non-integrable Hamiltonian $ H$,
while time-evolution on the right-hand side is with respect to the integrable $ H_0$, with $V^0(s) = e^{\ri  H_0 s}V e^{-\ri  H_0 s}$; see the Supplemental Material (SM).

A key feature of this perturbative approach is that it can describe
both the rapid onset of pre-thermalization and the slower process of
thermalization. As pre-thermalization builds up on a
$\lambda$-independent timescale,
 the state changes
abruptly but the conserved densities only receive small corrections
of order $\lambda^2$, as follows from
Eq.~\eqref{eqn_charge_general}. As a result the pre-thermalized state
is non-thermal, and is in fact close to a new GGE for the unperturbed
Hamiltonian $H_0$.  Afterwards, the dynamics occurs over timescales of
order $1/\lambda^2$. We will refer to this as the Boltzmann regime. It
is accessed by the formal $t\rightarrow\infty $ limit of
Eq.~\eqref{eqn_charge_general}, with $\b t = \lambda^2t$ held
fixed. In this limit, the unperturbed energy density is stationary,
while the $\b t$-derivatives of other observables take finite,
non-zero values, which satisfy the GGE equations of state. Proofs of these statements can be found in \cite{mallayya_prethermalization_2018,inprepa}. Thus in the
Boltzmann regime, the GGE continues to evolve slowly with time. The final
stationary regime is expected to occur for $t\gg 1/\lambda^2$, which
requires going beyond the perturbative result
\eqref{eqn_charge_general}; see \cite{lux_hydrodynamic_2014}. Nonetheless, for weakly broken
integrability, the Boltzmann regime is very long in comparison with
experimental timescales. Moreover, its physical properties are fully accessible using integrability as we now demonstrate.

{\em Form Factors.}--- As the right-hand side of
\eqref{eqn_charge_general} involves time-evolution under the
integrable Hamiltonian $H_0$, powerful techniques are available for
its evaluation.  The principal idea is that the matrix elements of the
perturbing operator $ v$ can be computed by means of a spectral
decomposition, in terms of a suitable basis of eigenstates of $H_0$.
For example, the initial GGE can be represented by a state
$\ket{{\rho_{\rm p}}}$, with $Z^{-1}\Tr (\rho \Or) =
\matrixel{\rho_{\rm p}}{\Or}{\rho_{\rm p}}$. Here, the quasiparticle
density $\rho_{\rm p}(\theta)$, as a function of the rapidity
$\theta$, is fixed by the thermodynamic Bethe ansatz (TBA)
\cite{Yang-Yang-1969,TakahashiTBAbook,ZAMOLODCHIKOV1990695}. Excited
states $\ket{\rho_{\rm p};\vec{p},\vec{h}}$ involve particle and hole
excitations on top of this
\cite{DfiniteTff2005,DfiniteTff2007,ChenDoyon2014,denardisParticle2018,CortesPanfil2019,CortesPanfil2019Gen},
where $\vec{p}$ and $\vec{h}$ indicate their respective sets of
rapidities. These diagonalize the momentum $Q_1$, energy $Q_2$, and other
conserved quantities $ Q_i$, with one-particle eigenvalues given by
$\kappa(\theta)$, $\varep(\theta)$ and $\eta_i(\theta)$ respectively.
Performing the spectral decomposition on
Eq.~\eqref{eqn_charge_general} yields
 \begin{equation}\label{expansion}
 	\partial_{\b t}\expectationvalue{q_i(0,t)}=2
 	\hspace{-2pt}\int\hspace{-2pt} \dd\tilde{\vec{p}}\dd\tilde{\vec{h}}\,\eta_i\delta(\kappa) \frc{\sin\varep t}{\varep} |\hspace{-2pt}\matrixel{\rho_{\rm p};\vec{p},\vec{h}}{v}{\rho_{\rm p}}\hspace{-2pt}|^2,
 \end{equation}
  as shown in the SM.  The integrand $\dd \tilde{\vec p} = \dd \vec p
 \rho_{\rm h}(\vec{p})$ includes the factor $\rho_{\rm
   h}(\vec{p})=\prod_{\theta\in\vec{p}}\rho_{\rm h}(\theta)$. This
 describes the accessible `phase space' given by the density of holes
 $\rho_{\rm h}(\theta)$, and likewise for $\rho_{\rm p}(\vec h)$ in terms
 of $\rho_{\rm p}(\theta)$. Here $\kappa=\sum_{\theta\in\vec{p}}
 \kappa(\theta) - \sum_{\gamma\in\vec{h}} \kappa(\gamma)$, and similarly
 for $\varep$ and $\eta_i$.

The expression \eqref{expansion} has a simple interpretation: in
accordance with \cite{dNBD,dNBD2}, particles and holes are in and out
states of scattering processes. The change in the charge density
$\expectationvalue{q_i(0,t)}$ is given by a weighted sum over all the
momentum conserving processes, with transition rates given by the form
factors squared $|\matrixel{\rho_{\rm p};\vec{p},\vec{h}}{v}{\rho_{\rm
    p}}|^2$ of the perturbing operator, in conformity 
with Fermi's golden rule. By evaluating these matrix elements one can
obtain a quantitative description of the thermalization process, from
short to long timescales.

{\em Pre-thermalization.}--- The form factor approach gives a quantitative
approach to pre-thermalization which is consistent with previous
results. For example, after an interaction quench, the charge
densities undergo fast initial dynamics, followed by an oscillatory
power-law approach to a quasi-stationary regime which persists for
long times. This can be verified by applying a small $\phi^4$
perturbation to a free massive scalar field, whose form factors can be
evaluated using the methods of \cite{ChenDoyon2014}. The results are provided in the SM;
similiar numerical results are obtained in \cite{berges_prethermalization_2004}.

{\em Boltzmann Regime.}--- After pre-thermalization, the approximate
GGE continues to evolve in accordance with
Eq.~\eqref{eqn_charge_general}. In the Boltzmann regime the
time-evolution of the state is slow, varying over long timescales of
order $1/\lambda^2$. As such, the change in the state can be large,
with the power-law tails describing the approach to the
instantaneous GGE giving perturbatively small corrections. In this regime, the
evolution is towards an (approximate) boosted thermal state for the
final Hamiltonian, in accordance with thermalization. Taking
$t\to\infty$, the evolution equations in this regime are given by
\beq\label{eqbolt} \partial_{\b t}\expectationvalue{q_i}_{\bm\beta(\b
  t)}= \int_{-\infty}^\infty \dd s \expectationvalue{[V^0(s),Q_i]
  v(0)}^{\rm c}_{\bm\beta(\b t)}, \eeq where the subscript ${\bm
  \beta}(\b t)$ indicates that the expectation value is taken in the
instantaneous GGE. As we demonstrate in the SM, a general H-theorem
shows that \eqref{eqbolt} is consistent with thermalization.

The spectral decomposition \eqref{expansion} available for integrable
systems allows us to recast \eqref{eqbolt} as a Boltzmann-type kinetic
equation. This sums over energy and momentum conserving scattering
processes with arbitrary numbers of particles.  This generalizes
approaches based on the kinetics of free models
\cite{boyanovsky_relaxation_1996,rau_reversible_1996,juchem_quantum_2004,benedettoFrom2008,hollands_derivation_2010,furst_derivation_2013,stark_kinetic_2013,drewes_boltzmann_2013,nessi_glass-like_2015,cauxnewton},
to interacting integrable systems.  Re-expressing \eqref{eqbolt} in
terms of the time-dependent quasiparticle density $\rho_{\rm
  p}(\theta)$, which represents the time-evolving GGE (see the SM), one
obtains
\begin{align}\label{expansionbolt}
	&\partial_{\b t}  \rho_{\rm p}(\theta) = I[\rho_{\rm p}](\theta) :=\\ &
	\int d\vec{p}d\vec{h}\,K(\theta,\vec{p})B(\vec{p}\rightarrow \vec{h})\big[\rho_{\rm h}(\vec{p})\rho_{\rm p}(\vec{h})-\rho_{\rm p}(\vec{p})\rho_{\rm h}(\vec{h})\big],\nonumber
\end{align}
where
\begin{equation}\label{scattering_kernel}
    B(\vec{p}\rightarrow \vec{h})=2\pi\delta(\kappa)\delta(\varepsilon)|\matrixel{\rho_{\rm p}}{v}{\rho_{\rm p};\vec{p},\vec{h}}|^2=B(\vec{h}\rightarrow \vec{p})
\end{equation}
is the matrix element for particle-hole scattering processes. In the special case of perturbations of free models, $K(\theta,\vec{p})=\sum_{\Phi\in\vec{p}}\delta(\theta-\Phi)$ and we have a generalization of the quantum Boltzmann equation to include higher-order scattering processes. If the unperturbed Hamiltonian $H_0$ is interacting, then $K$ also describes the effect of indirect processes where a particle of rapidity $\theta$ is created or destroyed in the interacting background in response to a scattering event. In this case
\begin{equation}
  K(\theta,\vec{p}) =\sum_{\Phi\in\vec{p}}K(\theta,\Phi),
\end{equation}
where
\begin{equation}
  K(\theta,\Phi) =\delta(\theta-\Phi)+\frc{\p}{\p\Phi}\left[\frc{F(\theta,\Phi)\rho_{p}(\Phi)}{\rho_p(\Phi)+\rho_h(\Phi)}\right].
\end{equation}
Here, $F(\theta,\Phi)$ is the backflow function representing the
effect of adding an excitation to the interacting background; see the
SM. Here we assume particle-hole symmetry, in accordance with the
usual microscopic reversibility condition of the Boltzmann scattering
kernel \eqref{scattering_kernel}. We show in the SM that an arbitrary boosted thermal state is a
fixed point of the time-evolution given in \eqref{expansionbolt},
confirming the general H-theorem presented there. We note finally that \eqref{expansionbolt} is an expansion in the number of excitations, which can often be recast as a low-density expansion. This is analogous to the LeClair-Mussardo series for equilibrium expectation values \cite{LM99}, which is observed to converge quickly.

\begin{figure}
    \centering
\includegraphics[width=\linewidth]{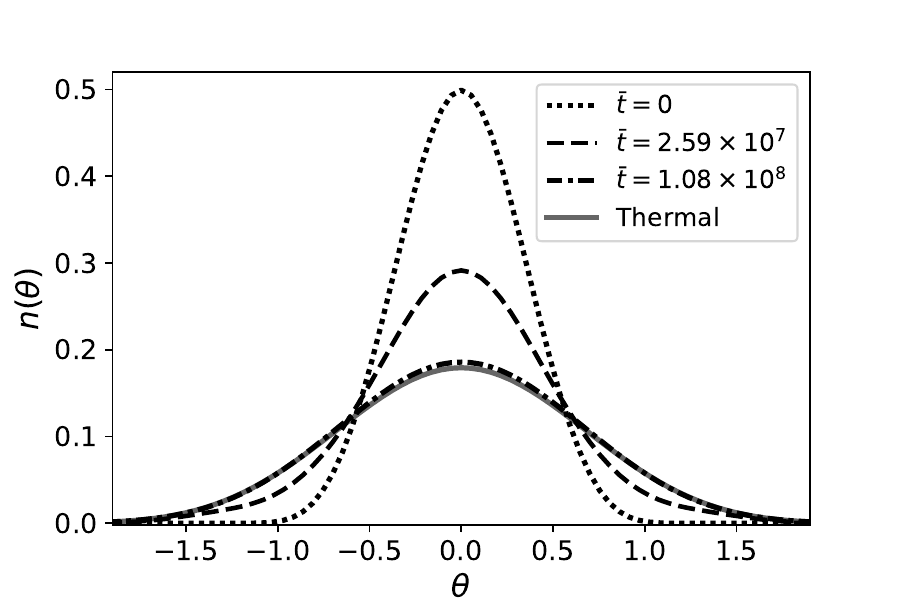}\\
    \includegraphics[width=\linewidth]{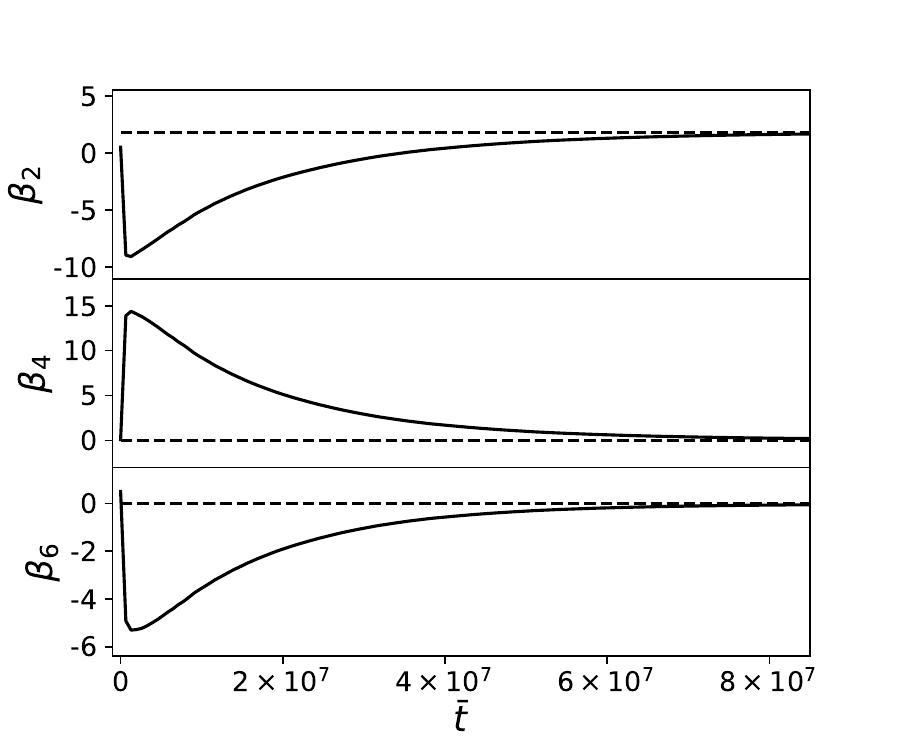}
    \caption{Upper panel: Time-evolution of the rapidity distribution $n(\theta)$ in the Boltzmann regime, for the free scalar field theory with unit mass perturbed by $\lambda/(6!)\int dx\,\phi^6(x)$, following a quantum quench from $\lambda=0$ to $\lambda> 0$. The initial distribution is the 
    GGE with $\beta_2$ = $\beta_6$ = $0.5$, $\beta_4$ = $0.1$, all other $\beta_{i}=0$. At late times $n(\theta)$ approaches a thermal distribution as indicated by the gray solid line. Lower panel: time-evolution of the first three effective inverse temperatures for the same quench, showing a non-monotonic  approach to thermalization. The large values of $\bar t$ reflects the standard normalization conventions for the scalar field theory, which effectively reduces the strength of
the $\phi^6$ perturbation. }
    \label{fig:Boltzmann}
\end{figure}

{\em Multiparticle Scattering.}--- The kinetic equation
\eqref{expansionbolt} generically contains infinitely-many scattering
processes with arbitrarily large numbers of particles $p\rightarrow h$. In the absence of internal degrees of freedom, the $2\rightarrow 2$ scattering processes do not contribute: these preserve momenta by 1+1-dimensional kinematics, hence the term in square brackets in \eqref{expansionbolt}
vanishes. This is consistent with the notion that thermalization
requires the non-trivial re-arrangement of momenta.  In
generic integrable models, the higher-particle form factors are
typically non-zero, thereby leading to thermalization via
\eqref{expansionbolt}. The $\phi^4$ theory considered above is
special, as these higher-particle form factors vanish. As
such, it does not thermalize in the Boltzmann regime in
$1+1$ dimensions, in agreement with three-loop results for correlation functions  \cite{berges_thermalization_2001,berges_quantum_2001}. For the $\phi^6$ perturbation, the $2\rightarrow 4$ and $3\rightarrow 3$ processes contribute. In Fig.~\ref{fig:Boltzmann} we show the time-evolution of the rapidity distribution $n(\theta)=2\pi\rho_{\rm p}(\theta)/\mathrm{cosh}(\theta)$, and the first few effective temperatures, in a $\phi^6$ quench. The results are consistent with thermalization, and illustrate how effective temperatures may exhibit non-monotonic dynamics.

In order to expose the relevant physics, we have concentrated for simplicity on a perturbation of a free model. However, an important aspect of this work is that it applies equally well to the case of an interacting integrable model. As an example, we consider the experimentally relevant case of two Lieb-Liniger gases perturbed by a density-density coupling. We consider arbitrary interaction strengths in the low-density regime. In this case, the two degrees of freedom allow for a non-trivial $2\rightarrow2$ contribution in the Boltzmann regime, for further details see the SM.

{\em Nearly-integrable Perturbations.}--- Perturbations that break integrability yet do not lead to thermalization in the Boltzmann regime can be seen as ``nearly-integrable perturbations", in analogy with the concept from KAM theory \cite{lazutkin_kam_1993}. The $\phi^4$ perturbation of the free massive scalar field discussed above is such an example. We show that such perturbations exist generically. To see this, we re-write the time-evolution
\eqref{eqbolt} as
\beq\label{boltdiff}
	\p_{\b t} \expectationvalue{q_i} = ([v,Q_i],v),
\eeq
where $(a,b)$ is a suitable inner product \cite{DoyonDiffusion2019}, defined by
\beq\label{innerproduct}
	(a,b)=\int \dd t\dd
x\,\expectationvalue{(1-\mathbb P)\big[a^0(x,t)\big]^\dag\, b(0,0)}^{\rm
  c}.
\eeq
Here, $a^0(x,t) = e^{\ri  H_0 t}a(x) e^{-\ri  H_0 t}$ and $\mathbb P$ is the projector onto the space of charges $Q_i$; see the SM. We show in the SM that current operators $j_k$, satisfying $\p_t q_k + \p_x j_k=0$, commute with the conserved charges under the inner product: $([j_k,Q_i],a)=0$ for all $a,i$. According to \eqref{boltdiff}, under a perturbation $v=j_k$, the state remains constant throughout the Boltzmann regime. Therefore {\em current operators are nearly-integrable perturbations}. This extends the notion of perturbed integrable models which preserve integrability in equilibrium \cite{zamo1989integrable,smirnov2017space,cavaglia2016TTbar,bargheer2009long,pozsgay2020TTbar,marchetto2020TTbar}. For example, there exist families of integrable models, $H = H_0 + V_\lambda$, which correspond to perturbations by current operators, $V_\lambda = \lambda \int \dd x\,j_k(x) + O(\lambda^2)$, at leading order  \cite{pozsgay2020current}. A similar relationship holds between the sine-Gordon model \cite{zamolodchikov1979factorized} and the $\phi^4$ perturbation of the scalar field. The observation here is that thermalization is absent at leading order, despite these models not being integrable.

The discussion above gives a natural classification of perturbations, and an associated classification of scattering processes. Indeed, under the inner product \eqref{innerproduct}, local operators form a Hilbert space $\mathcal H''$ \cite{DoyonDiffusion2019}. This admits an orthogonal decomposition $\mathcal H'' = \mathcal H_{\rm N} \oplus \mathcal H_{\rm B}$, where $\mathcal H_{\rm N}$ is the nearly-integrable subsector that commutes with $Q_i$ within $\mathcal H''$, and $\mathcal H_{\rm B}$ is the thermalizing Boltzmann subsector. 
In the kinetic description, operators in $\mathcal H_{\rm N}$ only couple to $2\rightarrow 2$ scattering processes. These, as explained above, do not lead to thermalization. It was shown in \cite{dNBD,dNBD2} that such processes lead to hydrodynamic diffusion instead, as they fully determine the Onsager matrix, $\mathcal L_{ij} = (j_i,j_j)$ \cite{Spohn-book,DoyonDiffusion2019}. Thus, there is a separation between processes leading to hydrodynamic diffusion, associated with $\mathcal H_{\rm N}$, and those leading to thermalization, associated with $\mathcal H_{\rm B}$.

{\em Thermalization and Entropy Production.}--- The late
time dynamics near the final, stationary state is obtained by
linearizing the evolution operator
\cite{spohn_phonon_2006}. In terms of the inverse effective temperatures
$\beta_i$, this gives \beq\label{boltmat} \sum_{j} \mathsf C_{ij}
\p_{\b t} \beta_j = -\sum_{j} \mathcal B_{ij}\beta_j; \eeq see the
SM. Here we define the {\em Boltzmann matrix} $\mathcal B_{ij} =
\big([v,Q_i],[v,Q_j]\big)$, while the static covariance matrix is $\mathsf
C_{ik} = \p \exv{q_i}/\p\beta_k$; both are non-negative and evaluated in the stationary
state. A similar evolution equation also holds for the small
deviations of the conserved densities $\delta q_i =
\exv{q_i}-\exv{q_i}_{\rm s}$. As $\mathcal B_{ij}=0$
for either $i,j=$ 1 or 2 the spectrum of $\Gamma=\mathcal B \mathsf C^{-1}$ always contains the eigenvalue 0, corresponding to the conserved modes of the Boltzmann
dynamics. The rest of the spectrum
controls the rate of approach to thermalization: if it extends
continuously to 0 then the approach is polynomial, whereas if there is a gap
of size $\gamma>0$,  it is exponential $\delta q_i\propto e^{- t / \tau}$ with $\tau = \lambda^{-2}\gamma^{-1}$  \cite{wennberg_stability_1995,herau_cauchy_2017,  spohn_phonon_2006}. It is notable that the timescale $\tau$ is solely determined by the final state, with the conserved energy and momentum densities containing the only information about the initial state.

\begin{figure}
    \centering
    \includegraphics[width=\linewidth]{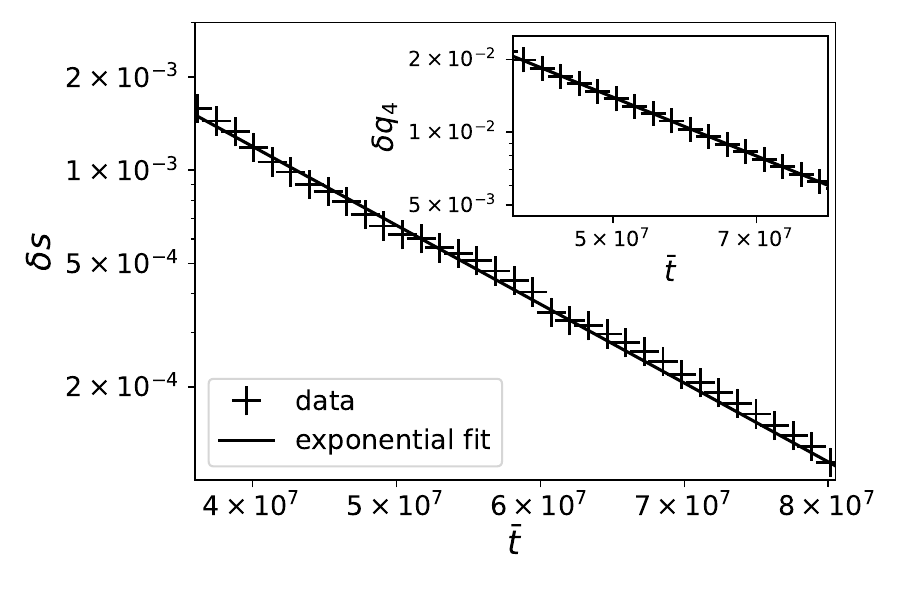}
    \caption{Exponential approach of the entropy and (inset) higher order charge $q_4$ to their stationary values, at times $\b t\gg 1$, for the same quench protocol as in Fig.~\ref{fig:Boltzmann}. The timescale $\gamma^{-1}\approx 3.58\times 10^7$ is found for $\delta q_4$, and $1.71\times 10^7$ for $\delta s$, in agreement with the theoretical value $\gamma^{-1}/2$.}
    \label{fig:approach_to_eq}
\end{figure}

In Fig.~\ref{fig:approach_to_eq}, we show numerical results consistent with an exponential approach to thermalization for the $\phi^6$ perturbation.  Therefore, for the $\phi^6$ perturbation, the spectrum of the Boltzmann matrix has a gap $\gamma>0$. At high temperatures we find an increasing thermalization timescale $\tau\sim T^\alpha$ with $\alpha\approx 3/2$, corresponding to an effectively gapless regime. In contrast, at low temperatures, we observe Arrenhius behavior with $\tau\sim e^{3m/T}$, corresponding to the 3-body collisions in the $\phi^6$ theory; see SM.

The Boltzmann matrix determines the late time dynamics of all physical quantities. Notably, the production of entropy near the final stationary state takes the form
\beq\label{entropy}
	\p_{\b t} s = \sum_{i,j\ge 3} \beta_i\mathcal B_{ij}\beta_j
	= ([v,\log\rho],[v,\log\rho]),
\eeq
where $\log\rho = -\sum_i \beta_i Q_i$ is the entropy operator; see the SM. As the right-hand side in \eqref{entropy} is quadratic in the $\beta_i$s, if there is a gap $\gamma$, the time-evolution of the entropy is also exponential, but with a rate $2\gamma$. This is twice that found in the time-evolution of the inverse temperatures and charge densities, which we confirm in Fig.~\ref{fig:approach_to_eq}.

Exponential decay can also be seen in correlation functions, as they are determined at large times by the conserved quantities. By projection methods, two-point functions at scaled wave numbers $\b k = k/\lambda^2 $ in the final state behave as 
\begin{multline}
\exv{ \Or_1\Or_2}^{\rm c}(\b k, \b t) =\\ -\sum_{ij} \partial_{\exv{q_i}}\hspace{-0.1cm}\exv{\Or_1} \exp\big[{\rm i}\mathsf A \b k \b t
	-\Gamma |\b t|\big]_{ij}\partial_{\beta_j}\hspace{-0.1cm}\exv{\Or_2}
\end{multline}	
where the matrix $\mathsf A_{ij} = \partial_{\exv{q_j}}\hspace{-0.1cm}\exv{j_i}$ encodes the propagation of the conserved modes, and $\Gamma$ their decay. In particular, this gives the Lorentzian broadening of the Drude peaks associated with the broken charges, $\int \dd \b t\, e^{\ri \b\omega \b t} \exv{ j_ij_k}^{\rm c}(\b k=0, \b t) = 2\big[\mathsf A (\Gamma^2+\b \omega^2)^{-1}\Gamma \mathsf A \mathsf C\big]_{ik}$, see also \cite{vas19}. We observe that the singularity in the complex $\omega$-plane that is nearest to the real line is at a distance $\gamma$. Dynamical correlation functions in the thermal state therefore determine the rate of approach towards it. A similar situation also occurs in holographic models, where the eigenvalues of the Boltzmann matrix are analogous to quasi-normal modes, see for example \cite{bhaseenHolographic2013}. As a signature of the integrability of the unperturbed model, this singularity is expected to be a branch point, because of the continuum of hydrodynamic modes parametrized by the rapidity $\theta$.

{\em Hydrodynamics.}--- The kinetic approach developed here is applicable beyond
quenches from homogeneous states, to include integrability destroying
perturbations in the hydrodynamic description of integrable models
\cite{PhysRevX.6.041065,PhysRevLett.117.207201}.  In this context, the effects of integrability breaking on the diffusive scale were recently discussed in \cite{vas19}. Here, we stress that the effects of weak perturbations are also manifest on the larger, Euler scale. In the Euler scaling
limit $x,t\to\infty$, $\lambda\to0$ with $\b t = \lambda^2 t$ and $\b
x = \lambda^2 x$ held fixed, the entropy increase of local fluid cells
occurs on Euler hydrodynamic timescales. The spectral
decomposition \eqref{expansionbolt} in the Boltzmann regime adds a
generalised collision term $I(\theta)$ to the fluid equations,
$\partial_{\b t}\rho_p(\theta)+\partial_{\b
  x}(v^{\mathrm{eff}}(\theta)\rho_p(\theta))= I(\theta)$, where
$v^{\rm eff}$ is given in
\cite{PhysRevX.6.041065,PhysRevLett.117.207201}. This opens the door to future studies of the crossover from integrable to non-integrable hydrodynamics, including the emergence of shocks, which are absent in the former case \cite{El-Kamchatnov-2005,bulint17,DDKYzeroentropy}.

{\em Conclusions.}--- In this work we have developed a form factor
approach to perturbed integrable models out of equilibrium. We have
shown that one can address a broad range of timescales, including the
approach to thermalization. We have provided analytical and numerical
predictions for the time-evolution of physical observables, including
conserved charges, effective temperatures, and rapidity distributions. We observe that the rate of thermalization for entropy is always exactly twice as large as that for conserved charges. We have also shown that there always exists a families of perturbations that do not thermalize in the weakly perturbed regime. It would be interesting to verify these predictions in experiment.

{\em Acknowledgments.}--- We are grateful to Jacopo De Nardis and Bal\'azs Pozsgay for comments and discussions. MJB acknowledges helpful conversations with Marcos Rigol, and BD thanks Wojciech De Roeck for enlightening discussions and Isabelle Bouchoule and J\'er\^ome Dubail for collaborations on related subjects. JD acknowledges funding from the EPSRC Centre for Doctoral  Training in Cross-Disciplinary Approaches to Non-Equilibrium Systems (CANES) under grant EP/L015854/1. BD acknowledges funding from the Royal Society under a Leverhulme Trust Senior Research Fellowship, ``Emergent hydrodynamics in integrable systems: non-equilibrium theory", ref.~SRF\textbackslash R1\textbackslash 180103. BD is grateful to the Tokyo Institute of Technology for support and hospitality during an invited professorship (October 2019), where part of this research was done. This research was further supported in part by the International Centre for Theoretical Sciences (ICTS) during a visit (JD, BD) for participating in the program Thermalization, Many body localization and Hydrodynamics (Code: ICTS/hydrodynamics2019/11).  All authors thank the Centre for Non-Equilibrium Science (CNES) and the Thomas Young Centre (TYC).


\begin{thebibliography}{99}
\bibitem{kinoshita}
 T.~Kinoshita, T.~Wenger, and D.~S. Weiss, {A Quantum Newton's Cradle}, {\em
   Nature}~{\bf 440}, 900~(2006). 
 
 \bibitem{PhysRevLett.98.050405}
 M.~Rigol, V.~Dunjko, V.~Yurovsky, and M.~Olshanii, {Relaxation in a
   Completely Integrable Many-Body Quantum System: An Ab Initio Study of the
   Dynamics of the Highly Excited States of 1D Lattice Hard-Core Bosons}, {\em
   Phys. Rev. Lett.}~{\bf 98}, 050405~(2007). 
 
 \bibitem{rigol_breakdown_2009}
 M.~Rigol, Breakdown of Thermalization in Finite One-Dimensional Systems, {\em Phys. Rev. Lett.}~{\bf 103}, 100403 (2009). 
 
 \bibitem{cassidy_generalized_2011}
 A. C. Cassidy, C. W. Clark, and M. Rigol, Generalized Thermalization in an Integrable Lattice System, {\em Phys. Rev. Lett.}~{\bf 106}, 140405 (2011). 
 
 \bibitem{gring_relaxation_2012}
 M. Gring, M. Kuhnert, T. Langen, T. Kitagawa et al., Relaxation and Prethermalization in an Isolated Quantum System, {\em Science}~{\bf 337}, 1318 (2012). 
 
 \bibitem{PhysRevLett.115.157201}
 E.~Ilievski, J.~De~Nardis, B.~Wouters, J.-S. Caux, F.~H.~L. Essler, and
   T.~Prosen, {Complete Generalized Gibbs Ensembles in an Interacting
   Theory}, {\em Phys. Rev. Lett.}~{\bf 115}, 157201~(2015). 
 
 \bibitem{Langen207}
 T.~Langen, S.~Erne, R.~Geiger, B.~Rauer et. al.
   {Experimental
   Observation of a Generalized Gibbs Ensemble}, {\em Science}~{\bf 348}, 207~(2015). 
 
 \bibitem{hydro1} P. Nozieres and D. Pines, {\em The Theory of Quantum Liquids} (Benjamin, New York, 1966).
 
 \bibitem{hydro2} S. Jeon and L. G. Yaffe, From Quantum Field Theory to Hydrodynamics: Transport Coefficients and Effective Kinetic Theory, {\em Phys. Rev. D}~{\bf 53}, 5799 (1996). 
 
 \bibitem{hydro3} A. G. Abanov, Hydrodynamics of correlated systems, in {\em Applications of Random Matrices in Physics}, edited by E. Br\'ezin, V. Kazakov, D. Serban, P. Wiegmann, and A. Zabrodin, NATO Science Series II: Mathematics, Physics and Chemistry, Vol. 221 (Springer, Dordrecht, 2006), pp. 139--161. 
 \bibitem{hydro4} E. Bettelheim, A. G. Abanov and P. Wiegmann, Nonlinear Quantum Shock Waves in Fractional Quantum Hall Edge States, {\em Phys. Rev. Lett.}~{\bf 97}, 246401 (2006). 
 
 \bibitem{BDLS} M. J. Bhaseen, B. Doyon, A. Lucas and K. Schalm, Far From Equilibrium Energy Flow in Quantum Critical Systems, {\em Nature Phys.}~{\bf 11}, 509~(2015). 
 
 \bibitem{PhysRevX.6.041065}
 O.~A. Castro-Alvaredo, B.~Doyon, and T.~Yoshimura, {Emergent Hydrodynamics in
   Integrable Quantum Systems Out of Equilibrium}, {\em Phys. Rev. X}~{\bf 6}, 041065~(2016). 
 
 \bibitem{PhysRevLett.117.207201}
 B.~Bertini, M.~Collura, J.~De~Nardis, and M.~Fagotti, {Transport in
   Out-of-Equilibrium $XXZ$ Chains: Exact Profiles of Charges and Currents},
   {\em Phys. Rev. Lett.}~{\bf 117}, 207201~(2016). 
 
 \bibitem{PhysRevB.97.045407}
 V.~B. Bulchandani, R.~Vasseur, C.~Karrasch, and J.~E. Moore, {Bethe-Boltzmann
   Hydrodynamics and Spin Transport in the XXZ Chain}, {\em Phys. Rev. B}~{\bf 97}, 045407~(2018). 
 
 \bibitem{cauxnewton}
 J.-S. Caux, B.~Doyon, J.~Dubail, R.~Konik, and T.~Yoshimura, {Hydrodynamics
   of the Interacting Bose Gas in the Quantum Newton Cradle Setup}, {\em SciPost Phys.}~{\bf 6}, 070 (2019). 
   
 \bibitem{Schemmerghd}
 M.~Schemmer, I.~Bouchoule, B.~Doyon, and J.~Dubail, {Generalized
   Hydrodynamics on an Atom Chip}, {\em Phys. Rev. Lett.}~{\bf 122}, 090601 (2019). 
 
 \bibitem{eisertreview}
 J.~Eisert, M.~Friesdorf, and C.~Gogolin, {Quantum Many-Body Systems Out of
   Equilibrium}, {\em Nature Phys.}~{\bf 11}, 124~(2015). 
 
 \bibitem{1742-5468-2016-6-064002}
 F.~H.~L. Essler and M.~Fagotti, {Quench Dynamics and Relaxation in Isolated
   Integrable Quantum Spin Chains}, {\em J. Stat. Mech. Theory Exp.}~{\bf 2016}, 064002~(2016). 
 
 \bibitem{VassMoo16}
 R. Vasseur and J. E. Moore, Nonequilibrium quantum dynamics and transport: from integrability to many-body localization, {\em J. Stat. Mech.}~{\bf 2016}, 064010 (2016). 
 
 \bibitem{IlievskietalQuasilocal} E. Ilievski, M. Medenjak, T. Prosen and L. Zadnik, Quasilocal Charges in Integrable Lattice Systems, {\em J. Stat. Mech.}~{\bf 2016}, 064008 (2016). 
 
 \bibitem{1742-5468-2016-6-064005}
 D.~Bernard and B.~Doyon, {Conformal Field Theory out of Equilibrium: a Review}, {\em {J. Stat. Mech. Theor. Exp.}}~2016, 064005~(2016). 
 
 \bibitem{DallesioETHreview2016}
 L. D'Alessio, Y. Kafri, A. Polkovnikov and M Rigol,
 From Quantum Chaos and Eigenstate Thermalization to Statistical Mechanics and Thermodynamics, {\em Adv. Phys.}~{\bf 65}, 239 (2016).
 
 
 \bibitem{GogolinEquilibration2016}
 C. Gogolin and J. Eisert, Equilibration, thermalisation, and the emergence of statistical mechanics in closed quantum systems, {\em Rep. Prog. Phys.}~{\bf 79}, 056001 (2016).
 
 
 \bibitem{DoyonLecture2020} B. Doyon, Lecture notes on generalised hydrodynamics, arXiv:1912.08496 (2019)
 
 \bibitem{lazutkin_kam_1993}
 V. F. Lazutkin, {\em KAM Theory and Semiclassical Approximations to Eigenfunctions}, 1st ed., A Series of Modern Surveys in Mathematics, Vol. 24 (Springer-Verlag, Berlin Heidelberg, 1993).
 
 \bibitem{mazets_breakdown_2008}
 I. E. Mazets, T. Schumm and J. Schmiedmayer, Breakdown of Integrability in a Quasi-1D Ultracold Bosonic Gas
 , {\em Phys. Rev. Lett.}~{\bf 100}, 210403 (2008).  
 
 \bibitem{moeckel_interaction_2008}
 M. Moeckel and S. Kehrein, Interaction Quench in the Hubbard Model, {\em Phys. Rev. Lett.}~{\bf 100}, 175702 (2008). 
 
 \bibitem{marcuzzi_prethermalization_2013}
 M. Marcuzzi, J. Marino, A. Gambassi, and A. Silva, Prethermalization in a Nonintegrable Quantum Spin Chain after a Quench, {\em Phys. Rev. Lett.}~{\bf 111}, 197203 (2013). 
 
 \bibitem{essler_quench_2014}
 F. H. L. Essler, S. Kehrein, S. R. Manmana and N. J.
 Robinson, Quench dynamics in a model with tuneable integrability breaking, {\em Phys. Rev. B}~{\bf 89}, 165104 (2014). 
 
 \bibitem{lux_hydrodynamic_2014}
 J. Lux, J. M\"uller,  A. Mitra and A. Rosch, Hydrodynamic long-time tails after a quantum quench, {\em Phys.  Rev. A}~{\bf 89}, 053608 (2014).
 
 
 \bibitem{mierzejewski_approximate_2015}
 M. Mierzejewski, T. Prosen and P. Prelov\v{s}ek, Approximate conservation laws in perturbed integrable lattice models, {\em Phys. Rev. B}~{\bf 92}, 195121 (2015). 
 
 \bibitem{bertini_prethermalization_2015}
 B. Bertini, F. H. L. Essler, S. Groha and N. J. Robinson, Prethermalization and Thermalization in Models with Weak Integrability Breaking, {\em Phys. Rev. Lett.}~{\bf 115}, 180601 (2015). 
 
 \bibitem{nessi_glass-like_2015}
 N. Nessi and A. Iucci, Glass-like Behavior in a System of One Dimensional Fermions after a Quantum Quench, arXiv:arXiv:1503.02507 (2015).
 
 \bibitem{bertini_thermalization_2016}
 B. Bertini, F. H. L. Essler, S. Groha and N. J. Robinson, Thermalization and light cones in a model with weak integrability breaking, {\em Phys. Rev. B}~{\bf 94}, 245117 (2016). 
 
 \bibitem{biebl_thermalization_2017}
 F. R. A. Biebl and S. Kehrein, Thermalization rates in the one-dimensional Hubbard model with next-to-nearest neighbor hopping, {\em Phys. Rev. B}~{\bf 95}, 104304 (2017). 
 
 \bibitem{albaPrethermalization2017}
 V. Alba and M. Fagotti, Prethermalization at Low Temperature: the Scent of Long-Range Order, {\em Phys. Rev. Lett.}~{\bf 119}, 010601 (2017). 
 
 \bibitem{lange_time-dependent_2018}
 F. Lange, Z. Lenar\v{c}i\v{c} and A. Rosch, Time-dependent generalized Gibbs ensembles in open quantum systems, {\em Phys. Rev. B}~{\bf 97}, 165138 (2018). 
 
 \bibitem{van_regemortel_prethermalization_2018}
 M. Van Regemortel, H. Kurkjian, I. Carusotto and
 M. Wouters, Prethermalization to thermalization crossover in a dilute Bose gas following an interaction ramp, {\em Phys. Rev. A}~{\bf 98}, 053612 (2018). 
 
 \bibitem{mallayya_prethermalization_2018}
 K. Mallayya, M. Rigol and W. De Roeck, Prethermalization and Thermalization in Isolated Quantum Systems, {\em Phys. Rev. X}~{\bf 9}, 021027 (2019). 
 
 \bibitem{tang_thermalization_2018}
 Y. Tang, W. Kao, K.-Y. Li, S. Seo et al., Thermalization near Integrability in a Dipolar Quantum Newton's Cradle, {\em Phys. Rev. X}~{\bf 8}, 021030 (2018). 
 
 \bibitem{vas19}
A. Friedman, S. Gopalakrishnan and R.Vasseur, {Diffusive hydrodynamics from integrability breaking}, 	arXiv:1912.08826 (2019)

\bibitem{Mori_thermalization_2018}
 T. Mori, T. N. Ikeda, E. Kaminishi and M. Ueda, Thermalization and prethermalization in isolated quantum systems: a theoretical overview, {\em J. Phys. B: At. Mol. Opt. Phys.}~{\bf 51}, 112001 (2018)
 
 \bibitem{mitra_quantum_2018}
 A. Mitra, Quantum Quench Dynamics, {\em Annual Review of Condensed Matter Physics}~{\bf 9}, 245 (2018).
 
 \bibitem{rieraThermalization2012}
 A. Riera, C. Gogolin and J. Eisert,  Thermalization in nature and on a quantum computer, {\em Phys. Rev.
 Lett.}~{\bf 108}, 080402 (2012).
 
 \bibitem{sirkerLocality2014}
 J. Sirker, N.P. Konstantinidis, F. Andraschko and N. Sedlmayr, Locality and thermalization in closed quantum systems, {\em Phys. Rev. A}~{\bf 89}, 042104 (2014).
 
 \bibitem{muellerThermalization2015}
 M.P. Mueller, E. Adlam, L. Masanes, N. Wiebe, Thermalization and canonical typicality in translation-
 invariant quantum lattice systems, {\em Commun. Math. Phys.}~{\bf 340}, 499 (2015).
 
 \bibitem{Doyon2017}
 B.~Doyon, {Thermalization and Pseudolocality in Extended Quantum Systems},  {\em Commun. Math. Phys.}~{\bf 351}, 155~(2017). 
 
 \bibitem{inprepa}
 Work in preparation.
 
 \bibitem{Yang-Yang-1969}
 C.~N. Yang and C.~P. Yang, {Thermodynamics of a One-Dimensional System of
   Bosons with Repulsive Delta-Function Interaction}, {\em J. Math. Phys.}~{\bf 10}, 1115~(1969). 
 
 \bibitem{TakahashiTBAbook} M. Takahashi, {\em Thermodynamics of One-Dimensional Solvable Models} (Cambridge University Press, Cambridge, 1999). 
 
 \bibitem{ZAMOLODCHIKOV1990695}
 A.~Zamolodchikov, {Thermodynamic Bethe Ansatz in Relativistic Models: Scaling
   3-State Potts and Lee-Yang Models}, {\em Nucl. Phys. B}~{\bf 342},
   695~(1990). 
   
 \bibitem{DfiniteTff2005} B. Doyon, Finite-temperature form factors in the Majorana theory, {\em J. Stat. Mech.}~{\bf 2005}, P11006 (2005).
 
 \bibitem{DfiniteTff2007} B. Doyon, Finite-temperature form factors: a review, {\em SIGMA}~{\bf 3}, 011 (2007).
 
 \bibitem{ChenDoyon2014} Y. Chen and B. Doyon, Form Factors in Equilibrium and Non-Equilibrium Mixed States of the Ising Model, {\em J. Stat. Mech. Theory Exp.}~{\bf 2014}, P09021 (2014)
 
 \bibitem{denardisParticle2018}
 J. De Nardis and M. Panfil, Particle-hole pairs and density-density correlations in the Lieb-Liniger model, {\em J. Stat. Mech.}~{\bf 2018}, 033102 (2018). 
 
 \bibitem{CortesPanfil2019}
 A. Cort\'es Cubero and M. Panfil, Thermodynamic bootstrap program for integrable QFT's: Form factors and correlation functions at finite energy density, {\em JHEP} 104 (2019). 
 
 \bibitem{CortesPanfil2019Gen}
 A. Cort\'es Cubero and M. Panfil, Generalized hydrodynamics regime from the thermodynamic bootstrap program, {\em SciPost Phys.}~{\bf 8}, 004 (2020). 
 
 \bibitem{dNBD} J. De Nardis, D. Bernard and B. Doyon, Hydrodynamic Diffusion in Integrable Systems, {\em Phys. Rev. Lett.}~{\bf 121}, 160603~(2018). 
 
 \bibitem{dNBD2} J. De Nardis, D. Bernard and B. Doyon, Diffusion in generalized hydrodynamics and quasiparticle scattering, {\em SciPost Phys.}~{\bf 6}, 049 (2019). 
 
 \bibitem{berges_prethermalization_2004}
 J. Berges, S. Bors\'anyi, and C. Wetterich, Prethermalization, {\em Phys. Rev. Lett.}~{\bf 93}, 142002 (2004). 
 


 \bibitem{boyanovsky_relaxation_1996}
 D. Boyanovsky, I. D. Lawrie and D. S. Lee, Relaxation and kinetics in scalar field theories, {\em Phys. Rev. D}~{\bf 54}, 4013 (1996). 
 
 \bibitem{rau_reversible_1996}
 J. Rau and B. M\"uller, From reversible quantum microdynamics to irreversible quantum transport, {\em Phys. Rep.}~{\bf 272}, 1 (1996). 
 
 \bibitem{juchem_quantum_2004}
 S. Juchem, W. Cassing and C. Greiner, Quantum dynamics and thermalization for out-of-equilibrium $\varphi^4$ theory, {\em Phys. Rev. D}~{\bf 69}, 025006 (2004). 
 
 \bibitem{benedettoFrom2008}
 D. Benedetto, F. Castella, R. Esposito, and M. Pulvirenti, From the N-body Schrödinger Equation to the Quantum Boltzmann Equation: a Term-by-Term Convergence Result in the Weak Coupling Regime, {\em Commun. Math. Phys.}~{\bf 277}, 1 (2008). 
 
 \bibitem{hollands_derivation_2010}
 S. Hollands and G. Leiler, On the derivation of the Boltzmann equation in quantum field theory: Flat spacetime, arXiv:1003.1621 (2010)
 
 \bibitem{furst_derivation_2013}
 M. L. R. F\"urst, J. Lukkarinen, P. Mei and H. Spohn, Derivation of a matrix-valued Boltzmann equation for the Hubbard model,  {\em J. Phys. A: Math. Theor.}~{\bf 46}, 485002 (2013). 
 
 \bibitem{stark_kinetic_2013}
 M. Stark and M. Kollar, Kinetic description of thermalization dynamics in weakly interacting quantum systems, arXiv:1308.1610 (2013)
 
 \bibitem{drewes_boltzmann_2013}
 M. Drewes, S. Mendizabal and C. Weniger, The Boltzmann equation from quantum field theory, {\em Phys. Lett. B}~{\bf 718}, 1119 (2013). 

\bibitem{LM99} A. LeClair and G. Mussardo, Finite temperature correlation functions in integrable QFT, {\em Nuclear Physics B} {\bf 552}, 624 (1999).

 \bibitem{berges_thermalization_2001}
 J. Berges and J. Cox, Thermalization of quantum fields from time-reversal invariant evolution equations, {\em Phys. Lett. B}~{\bf 517}, 369 (2001).
 
 
 \bibitem{berges_quantum_2001}
 J. Berges, Quantum Fields far from Equilibrium and Thermalization, {\em Non-Perturbative QCD, Proc. of the Sixth Workshop},  Ed. H. M. Fried, Y. Gabellini and B. M\"uller, 85 (2002).
 

\bibitem{DoyonDiffusion2019}
B. Doyon, Diffusion and superdiffusion from hydrodynamic projections, preprint arXiv:1912.01551 (2019).

\bibitem{zamo1989integrable}
A. B. Zamolodchikov, Integrable Field Theory from Conformal Field Theory, {\em  Adv. Stud. Pure Math.} {\bf 19}, 641 (1989).

\bibitem{smirnov2017space}
F. A. Smirnov and A. B. Zamolodchikov, On space of integrable quantum field theories, {\em Nucl. Phys. B} {\bf 915}, 363 (2017).

\bibitem{cavaglia2016TTbar} A. Cavagli\`a, S. Negro, I. M. Sz\'ecs\'enyi, and R. Tateo, $T\b{T}$ deformed 2D Quantum Field Theories, {\em JHEP} {\bf 10}, 112 (2016).

\bibitem{Spohn-book}
 H.~Spohn, {\em {Large Scale Dynamics of Interacting Particles}} (Springer-Verlag, Heidelberg, 1991)

\bibitem{bargheer2009long}
T. Bargheer, N. Beisert and F. Loebbert, Long-Range Deformations for Integrable Spin Chains, {\em J. Phys. A} {\bf 42}, 285205 (2009).

\bibitem{pozsgay2020TTbar}
B. Pozsgay, Y. Jiang and G. Tak\'acs, $T\b{T}$-deformation and long range spin chains, {\em JHEP} {\bf 03}, 092 (2020).

\bibitem{marchetto2020TTbar}
E. Marchetto, A. Sfondrini and Z. Yang, $T\b{T}$-deformation and integrable spin chains, {\em Phys. Rev. Lett.} {\bf 124}, 100601 (2020).

\bibitem{pozsgay2020current}
B. Pozsgay, Current operators in integrable spin chains: lessons from long range deformations, {\em SciPost Phys.} {\bf 8}, 016 (2020).
 
\bibitem{zamolodchikov1979factorized}
A.  B.  Zamolodchikov  and  Al.  B.  Zamolodchikov,  Factorized  S-matrices  in  two-dimensions as the exact solutions of certain relativistic quantum field models, {\em Annals Phys.} {\bf 120}, 253 (1979).

 \bibitem{spohn_phonon_2006}
 H. Spohn, The Phonon Boltzmann Equation, Properties and Link to Weakly Anharmonic Lattice Dynamics, {\em J. Stat. Phys.}~{\bf 124}, 1041 (2006). 
 
 \bibitem{wennberg_stability_1995}
 B. Wennberg, Stability and exponential convergence for the Boltzmann equation, {\em Arch. Rational Mech. Anal.}~{\bf 130}, 103 (1995). 
 
 \bibitem{herau_cauchy_2017}
 F. H\'erau, D. Tonon and I. Tristani, Regularization estimates and Cauchy theory for inhomogeneous Boltzmann equation for hard potentials without cut-off, arXiv:1710.01098 (2017).
 
 \bibitem{bhaseenHolographic2013}
 M. J. Bhaseen, J. P. Gauntlett, B. D. Simons, J. Sonner and T. Wiseman, Holographic Superfluids and the Dynamics of Symmetry Breaking,
 {\em Phys. Rev. Lett.}~{\bf 110}, 015301 (2013). 
 
 \bibitem{DDKYzeroentropy}  B. Doyon, J. Dubail, R. Konik and T. Yoshimura, Large-scale description of interacting one-dimensional Bose gases: Generalized hydrodynamics supersedes conventional hydrodynamics, {\em Phys. Rev. Lett.}~{\bf 119}, 195301 (2017). 
 
 \bibitem{El-Kamchatnov-2005}
 G.~A. El and A.~Kamchatnov, {Kinetic Equation for a Dense Soliton Gas}, {\em
   Phys. Rev. Lett.}~{\bf 95}, 204101~(2005). 
 
 \bibitem{bulint17}
 V. B. Bulchandani, On classical integrability of the hydrodynamics of quantum integrable systems, {\em J. Phys. A: Math. Theor.}~{\bf 50}, 435203 (2017). 
 
 \end{thebibliography}
\end{document}


\bc\Large Supplemental Material for\\ ``Non-Equilibrium Dynamics and Weakly Broken integrability"\ec
\bc
Joseph Durnin, M. J. Bhaseen, Benjamin Doyon\ec
\section{Time-Evolution of the Charge Densities}\label{secpert}
We here derive a perturbative equation of motion for the densities of conserved charges, showing in particular that the leading term occurs in second order perturbation theory. We consider the Hamiltonian $H=H_0+\lambda V$, composed of an integrable contribution $H_0$ and a homogeneous integrability-breaking perturbation $V= \int \dd x\,v(x)$. The conserved charges $Q_i=\int \dd x\,q_i(x)$ of the unperturbed Hamiltonian satisfy $[H_0,Q_i]=0$. We further assume that they mutually commute, $[Q_i,Q_j]=0$ for all $i,j$, and define $Q_1=P$ and $Q_2=H_0$ as in the main text. Expanding the time-evolved charge densities $q_i(x,t) = e^{\ri H t} q_i(x) e^{-\ri H t}$ to first order in $\lambda$, perturbation theory gives
\begin{equation}
    q_i(x,t)=q_i^0(x,t)+\ri \lambda\int_0^t\dd s\,[V^0(s),q_i^0(x,t)]+O(\lambda^2),
\end{equation}
where the superscript $0$ indicates time-evolution with respect to $H_0$, i.e.~$q_i^0(x,t) = e^{\ri H_0 t} q_i(x) e^{-\ri H_0 t}$ and $V^0(s) = e^{\ri H_0 s} V e^{-\ri H_0 s}$. We evaluate this in the initial state with density matrix $\rho_0 =Z^{-1} e^{-\sum_i \beta^i Q_i}$
\begin{equation} \label{stationarity_eq}
    [\rho_0,Q_j]=0 \;\forall\,j.
\end{equation}
This follows from the identity $[Q_i,Q_j]=0$; taking $j=1$ implies spatial homogeneity of the state, and $j\ne 1$ implies invariance of the state under the action of the associated charge $Q_j$. The Heisenberg equation of motion $\partial_tq_i(x,t)=\ri [H,q_i(x,t)]$ then gives
\begin{equation}\label{pertgen}
    \partial_t\exv{q_i(0,t)}=\ri\lambda\exv{\left[V,q_i^0(0,t)\right]}-\lambda^2\int_0^t\dd s\exv{\left[V,\left[V^0(s),q_i^0(0,t)\right]\right]}+O(\lambda^3),
\end{equation}
where spatial homogeneity allows evaluation at $x=0$ without loss of generality. The first term on the right-hand side can be shown to vanish by repeated application of Eq.~\eqref{stationarity_eq} for different values of $j$:
\begin{equation}\label{vanish}
 \exv{\left[V,q_i^0(0,t)\right]} = \int dx\expectationvalue{\left[v(x),q_i^0(0,t)\right]}=\int dx\expectationvalue{\left[v,q_i^0(-x,t)\right]}=\expectationvalue{\left[v,Q_i\right]}=0.
\end{equation}
Similar manipulations allow the second term to be written in a more convenient form
\begin{align} 
    \partial_t\expectationvalue{q_i(0,t)}&=-\lambda^2\int \dd x\int_0^t\dd s\expectationvalue{\left[v(0),\left[v^0(x,s),Q_i\right]\right]}
    \nonumber \\&=\lambda^2\int \dd x\int_{-t}^t\dd s\expectationvalue{\left[v^0(x,s),Q_i\right]v(0)}^{\rm c}
    \nonumber \\&=\lambda^2 \int_{-t}^t\dd s\expectationvalue{\left[V^0(s),Q_i\right]v(0)}^{\rm c},\label{eqn_golden_rule}
\end{align}
which corresponds to Eq.~(1) in the main text. In the second line we have replaced the correlation function with its connected part,
\beq
	\expectationvalue{\left[v^0(x,s),Q_i\right]v(0)}^{\rm c}
	=
	\expectationvalue{\left[v^0(x,s),Q_i\right]v(0)}-
	\expectationvalue{\left[v^0(x,s),Q_i\right]}
	\expectationvalue{v(0)},
\eeq
which is possible as $\expectationvalue{\left[v^0(x,s),Q_i\right]}\expectationvalue{v(0)}=0$ by Eq.~\eqref{stationarity_eq}. This makes the convergence of the integral in $x$ immediate by the clustering properties of the state.
\section{$\phi^p$ Perturbation of the Free Scalar Field} \label{sectPhip}
In this section we provide further details on the results presented in the main text for pre-thermalization and thermalization in a weakly-interacting bosonic field.
For $\phi^p$ perturbations of the free scalar field the form factor expansion for the charge evolution truncates at finite order; for a polynomial interaction $\phi^p$ the form factors with more than $p$ particles/holes are vanishing. We consider the unperturbed Hamiltonian
\begin{equation}
    H_0=\frc12 \int \dd x :\left[\pi^2+\left(\nabla \phi\right)^2+\phi^2 \right]:,
\end{equation}
where the normal ordering $:\Or:$ is made with respect to the vacuum, and for simplicity we set the mass to 1. The Bose distribution functions associated with the GGE density matrix are
\begin{equation}
    n(\theta)=\frac{1}{e^{W(\theta)}-1}\;,\;\;\bar{n}(\theta)=1+n(\theta),
\end{equation}
where
\begin{equation} \label{w_eqn}
    W(\theta)=\sum_{n=1}^\infty\left(\beta_{2n-1}\mathrm{sinh}(n\theta)+\beta_{2n}\mathrm{cosh}(n\theta)\right).
\end{equation}
The quasi-particle density is $\rho_{\rm p}(\theta) = \cosh(\theta)n(\theta)/(2\pi)$, as given in the main text.
The form factor expansion for correlation functions can be obtained by the methods of Ref.~\cite{DfiniteTff2005,DfiniteTff2007,ChenDoyon2014}, by application of Wick's theorem for bosonic commutation relations. Taking the perturbation $V=\lambda/(4!)\int \dd x\,\phi^4(x)$, only the 2 and 4 particle terms contribute to the equation of motion for the charge densities. For the numerical data presented the 4-particle terms were dominant. This yields the following evolution equation in the prethermal regime
\begin{figure}[p]
    \centering
    \includegraphics[width=0.75\linewidth]{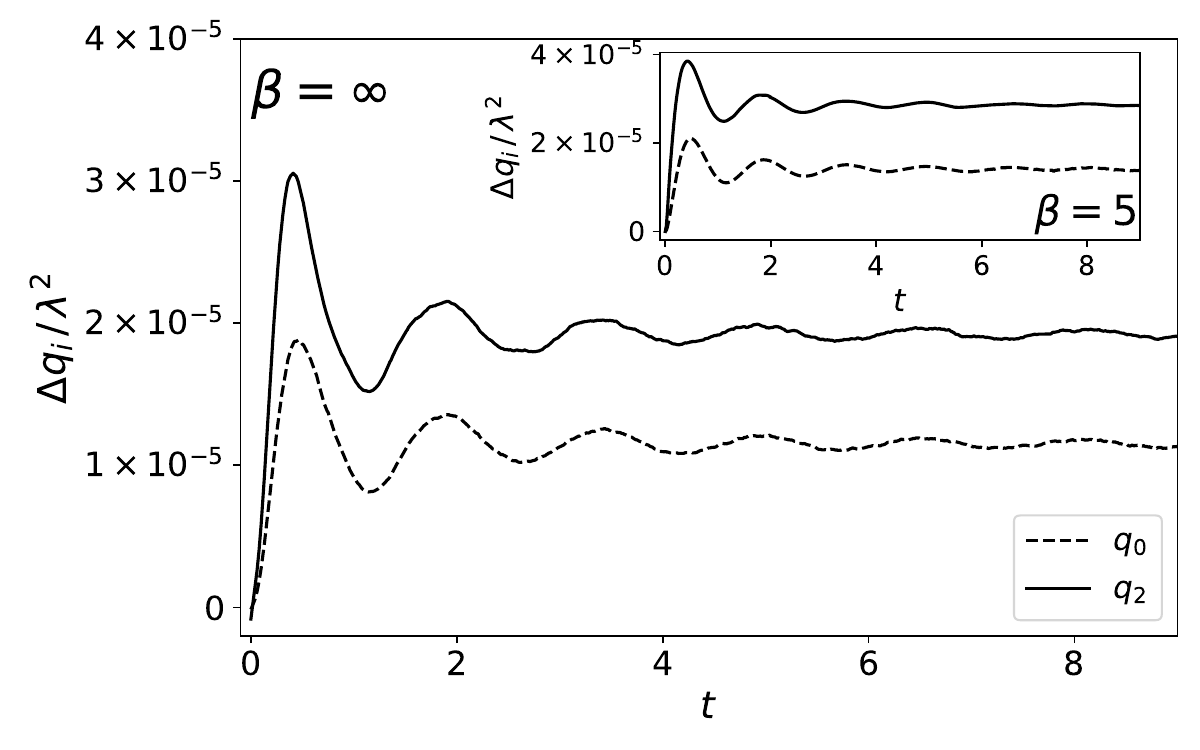}
\caption{Change in the particle and energy densities $q_0$ and $q_2$, respectively, for the free
  scalar field theory (with unit mass) perturbed by
  $\lambda/(4!)\int dx\,\phi^4(x)$, following a quantum quench from $\lambda=0$ to
  $\lambda> 0$. The initial state is taken as the vacuum state
  (main panel), and a thermal state with $\beta=5$ (inset). In both
  cases, the charge densities settle down to constant values corresponding to pre-thermalization.}
    \label{fig:prethermal}
\end{figure}
\begin{align} 
    \partial_t\exv{q_i(0,t)}=&\frac{2}{4!}\left(\frac{\lambda}{2^4\pi^2}\right)^2\int \dd\theta_1\dd\theta_2\dd\theta_3\dd\theta_4\,\delta(\kappa)\frac{\sin(\varepsilon t)}{\varepsilon}\nonumber 
    \\
    &
    \big[\bar{n}_1\bar{n}_2\bar{n}_3\bar{n}_4\,(h_{i}(\theta_1)+h_{i}(\theta_2)+h_{i}(\theta_3)+h_{i}(\theta_4))\nonumber 
    \\
    &+16\bar{n}_1\bar{n}_2\bar{n}_3n_4\,(h_{i}(\theta_1)+h_{i}(\theta_2)+h_{i}(\theta_3)-h_{i}(\theta_4))\nonumber 
    \\
    &+36\bar{n}_1\bar{n}_2n_3n_4\,(h_{i}(\theta_1)+h_{i}(\theta_2)-h_{i}(\theta_3)-h_{i}(\theta_4)) \nonumber 
    \\
    &
    +16\bar{n}_1n_2n_3n_4\,(h_{i}(\theta_1)-h_{i}(\theta_2)-h_{i}(\theta_3)-h_{i}(\theta_4))\nonumber
    \\
    &+n_1n_2n_3n_4\,(-h_{i}(\theta_1)-h_{i}(\theta_2)-h_{i}(\theta_3)-h_{i}(\theta_4))
    \big], \label{phi_4_pretherm}
\end{align}
where $n_i = n(\theta_i)$ and $\b n_i = \b n(\theta_i)$. As shown in Figs.~\ref{fig:prethermal} and \ref{fig:oscillations}, pre-thermalization in the $\phi^4$ theory is seen to consist of a fast increase in the value of charge densities, followed by oscillations within a decaying envelope to the long-time pre-thermal value. Numerically, the envelope is found to decay as $t^{-\alpha}$ with $\alpha = 1.0\pm0.2$ for all observables and initial states studied; see Fig.~\ref{fig:oscillations} for an illustration. The numerical solution of Eq.~\eqref{phi_4_pretherm} was obtained using the numerical integration routines in Mathematica.
\begin{figure}[p]
    \centering
    \includegraphics[width=0.48\linewidth]{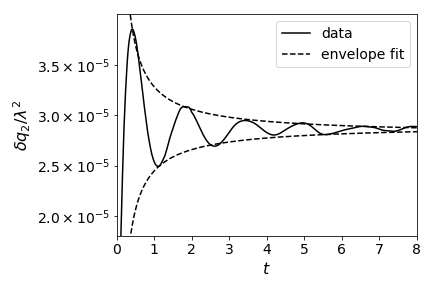}
    \includegraphics[width=0.48\linewidth]{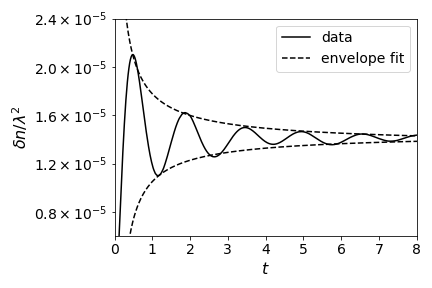}
    \caption{Time-evolution of the particle and energy densities $n=q_0$ and $q_2$ after a $\lambda/(4!)\phi^4$ quench of a free scalar field with unit mass, with a thermal initial state at inverse temperature $\beta = 5$. Oscillations are observed within an envelope that is well described by a power law $t^{-1}$.
    }
    \label{fig:oscillations}
\end{figure}

\begin{figure}
    \hspace{0.1\linewidth}(\bf{a})\hspace{0.44\linewidth}(b)\\
    \includegraphics[width=0.48\linewidth]{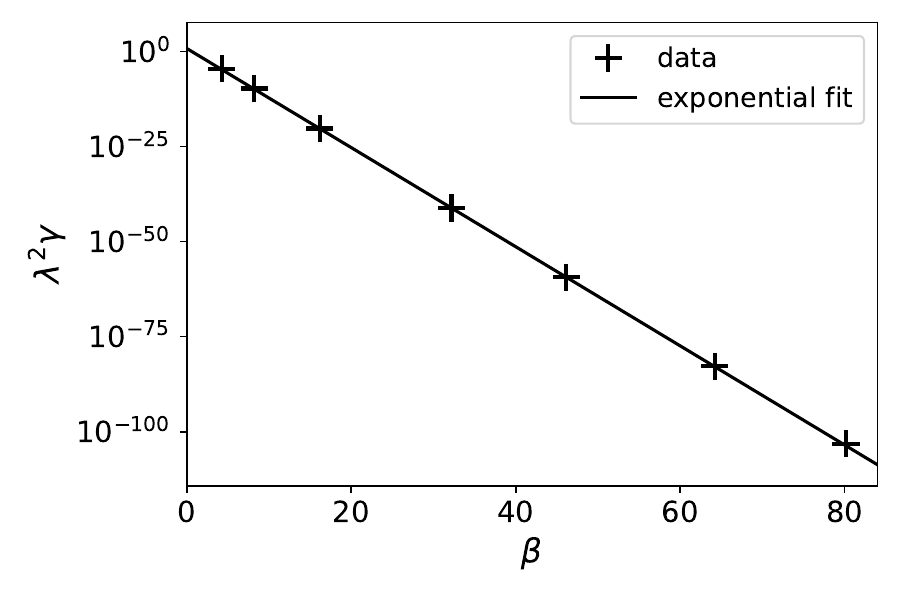}
    \includegraphics[width=0.48\linewidth]{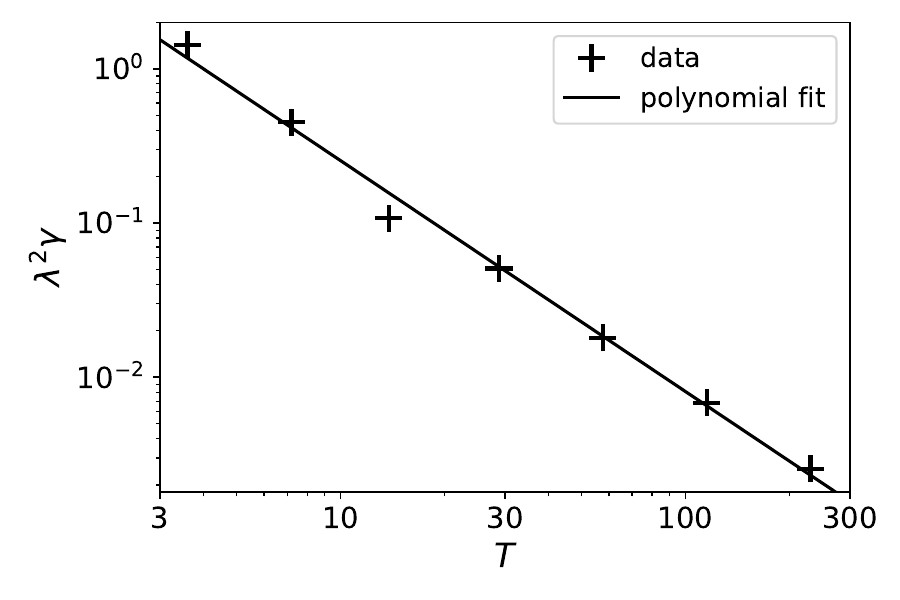}
    \caption{({\bf a}): Exponential decay of the rate of approach to thermal equilibrium at low temperatures $\beta\gg 1$, showing the relationship $\lambda^2\gamma\sim e^{-3\beta}$. ({\bf b}): Polynomial decay of the rate of approach to thermal equilibrium at high temperatures $T\gg 1$, showing the relationship $\lambda^2\gamma\sim T^{-\alpha}$ where $\alpha = 1.498\pm 0.003$.}
    \label{fig:rate_T_dependence}
\end{figure}

In order to go beyond pre-thermalization, we turn our attention to the $\phi^6$ theory with $V=\lambda/(6!)\int \dd x\,\phi^6(x)$. As we discuss in the main text, this has a non-trivial Boltzmann regime. Here the only terms which contribute to the kinetic equation occur at the 6-particle/hole level
\begin{align}\label{kinetic_eq}
    \partial_{\b t}n(\theta)=&\frac{1}{6!\cosh{\theta}}\left(\frac{1}{2^5\pi^2}\right)^2\int \dd\theta_1\dd\theta_2\dd\theta_3\dd\theta_4\dd\theta_5\delta(\kappa)\delta(\varepsilon)\nonumber\\
    \bigg[&1200\left(\bar{n}(\theta)\bar{n}_1\bar{n}_2n_3n_4n_5-n(\theta) n_1n_2\bar{n}_3\bar{n}_4\bar{n}_5\right) \nonumber \\
    &900\left(\bar{n}(\theta)\bar{n}_1\bar{n}_2\bar{n}_3n_4n_5-n(\theta) n_1n_2n_3\bar{n}_4\bar{n}_5\right) \nonumber \\
    &450\left(\bar{n}(\theta)\bar{n}_1n_2n_3n_4n_5-n(\theta) n_1\bar{n}_2\bar{n}_3\bar{n}_4\bar{n}_5\right)\bigg].
\end{align}
Numerical solution of this equation gives the results presented in Figs (1) and (2) in the main text. As shown in Fig \ref{fig:rate_T_dependence}, it is also used to find the dependence of the rate of approach to equilibrium $\gamma$ on the temperature $T$ of the final state.

In order to solve the generalized Boltzmann equation \eqref{kinetic_eq} numerically, we implement the energy and momentum delta functions analytically, facilitated by the change of variables $k_i=\sinh(\theta_i)$. We take a discrete set of values of $\theta$ and interpolate between $n(\theta)$ at these points to obtain the full function $n(\theta)$, as fast decay of $n(\theta)$ at large $|\theta|$ (cf.~equation \eqref{w_eqn}) means that we can restrict the infinite integrals over rapidity to some suitable finite region $\Lambda$. We then solve using the forward Euler method in $t$, and evaluate the remaining triple integrals by Monte-Carlo methods. We estimate the errors at $\sim 0.1\%$ using the ratio of the standard deviation to the mean of the energy $\sigma_{q_2}/\bar{q}_2$ over the simulation run-time. Energy fluctuations occur as a result of the Monte-Carlo methods used, and linear interpolation between lattice points in evaluating integrals involving $n(\theta)$. Uncertainty in the fitting parameters is found to be of this order of magnitude, except for the entropy decay rate which is very sensitive to energy fluctuations (cf.~Eq.~\eqref{eqn_entropy}), and therefore the uncertainty is an order of magnitude larger.

\section{Coupled Lieb-Liniger Gases}\label{LL_Sect}
The main results of the paper apply to perturbations of interacting 
integrable systems as well as perturbations of free theories; the 
derivation is presented in Section \ref{Sec_spectral_expansion} below. In order to illustrate this, 
let us consider a perturbed Lieb-Liniger system with the Hamiltonian
\begin{equation}\label{LL_Hamilton}
    H=\int dx\,\left(-\sum_{i=a,b}\psi_i^\dagger\p^2\psi_i+\sum_{i=a,b}2c_i\psi^\dagger_i\psi^\dagger_i\psi_i\psi_i+\lambda\psi^\dagger_a\psi_a\psi^\dagger_b\psi_b\right).
\end{equation}
This is a case where the system contains multiple internal degrees of freedom. This can be accommodated in the main results by allowing the rapidity to carry an additional index denoting the relevant degree of freedom.

The system \eqref{LL_Hamilton} has two integrable points, at $\lambda=0$ where we have a two-species Lieb-Liniger model, and at $\lambda=4c_a=4c_b$ when it becomes the Yang-Gaudin model. We will consider the case of a small $\lambda$ perturbation of the former integrable point. As the system has multiple particle species, $Q_i$ carries a compound 
index $i$ describing both the order of the charge and the particle species. The spectral expansion is over product states $\ket{\Omega}=\ket{\Omega}_a\bigotimes\ket{\Omega}_b$. The relevant thermodynamic form factors, corresponding to the density operator in the Lieb-Liniger model, are known exactly \cite{denardis2015density}. 

In this model, the form factor expansion in excitation numbers
can be recast as a low-density expansion, in powers of the density $D=\int\, d\theta\rho_\mathrm{p}(\theta)$ of the gas. In particular, the phase space factors appearing 
in (4) of the main text are dominant for smaller numbers of 
particle-hole excitations. At low enough densities
the number of excitations is typically small, by analogy with the 
Leclair-Mussardo formula for equilibrium expectation values \cite{LM99}.

We consider the lowest order contribution in the density for arbitrary 
Lieb-Liniger couplings $c_i$. At this order, only the two particle-hole 
terms remain; these do not vanish due to the presence of multiple
particle species. The results are obtained by taking the zero-density 
limit of the density operator form factors.
Eliminating the energy and momentum delta functions, we find the following kinetic equation
\begin{figure}
    \centering
    \includegraphics[width=0.9\textwidth]{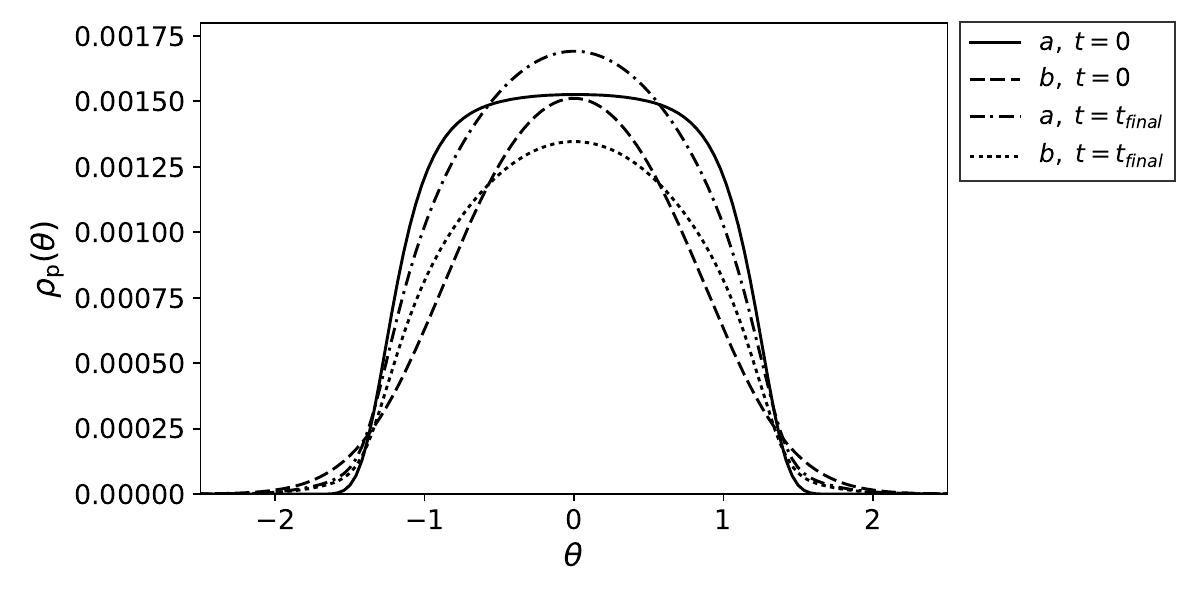}
    \caption{Time-evolution of the distribution functions $\rho_\mathrm{p}^i(\theta)$ in the Boltzmann regime of the evolution under the Hamiltonian \eqref{LL_Hamilton}, described by Eq.~\eqref{LL_eqmot}. Gas $a$ is initially in the GGE state with $\beta^a_i=\sum_{n=0}^4\delta_{2n,i}/10$, and gas $b$ in the thermal state with $\beta^b_0=1/10$, $\beta^b_2=13/10$.}
    \label{fig:LL_graph}
\end{figure}
\begin{align}\label{LL_eqmot}
    \p_{\bar{t}}\rho^a_p(p)=&\frac{(2\pi)^2}{16}\int \frac{dh}{|p-h|}\frac{\phi_a^2(p-h)\phi_b^2(p-h)}{(p-h)^4}\left((p-h)^2+c_a^2\right)\left((p-h)^2+c_b^2\right)\nonumber \\
    &\hspace{60pt}\times\left(\rho_\mathrm{p}^a(h)\rho_\mathrm{p}^b(p)-\rho_\mathrm{p}^a(p)\rho_\mathrm{p}^b(h)\right)\nonumber \\
    =&-\p_{\bar{t}}\rho^b_p(p),
\end{align}
where the scattering phase $\phi_i(p)=2\mathrm{arctan}(p/c_i)$. In this case the scattering only involves exchange of rapidities between 
the two gases, and indirect scattering is absent. We recall that 
indirect scattering processes are interpreted as processes whereby a 
particle of rapidity $\theta$ is created or removed by a change in the 
interacting background induced by scattering processes.

In Fig \ref{fig:LL_graph} we show the time-evolution of the distribution functions in the Boltzmann regime, where we take GGE initial conditions with $c_a=1$, $c_b=5$. The numerical solution to \eqref{LL_eqmot} was obtained in a similar way to \eqref{kinetic_eq}, as outlined in section \ref{sectPhip}. We observe an exponential rate of approach to the equilibrium value, 
where the rate of approach for the entropy is twice that for the 
charges, as predicted by the theory in the main text.
\section{Thermodynamic Spectral Expansion} \label{Sec_spectral_expansion}
We now provide more details on the spectral expansion leading to the scattering-like Eqs (2) and (4) in the main text. The right-hand side of equation \eqref{eqn_golden_rule} is a two-point function, allowing the spectral decomposition
\beq \label{spec1}
    \partial_{\lambda^2t}\exv{q_i(0,t)}=
    \int_{-t}^t \dd s \sum_{\Omega}
	\matrixel{\rho_{\rm p}}{[V^0(s),Q_i]}{\Omega}\matrixel{\Omega}{ v(0,0)}{\rho_{\rm p}},
\eeq
where $\Omega$ labels a complete set of states. The state $\ket{\rho_{\rm p}}$ can be given explicitly in free models \cite{DfiniteTff2005,DfiniteTff2007,ChenDoyon2014}, and using the equivalence of the microcanonical and canonical ensembles it can be interpreted in more general integrable systems as a representative Bethe state $\ket{\rho_{\rm p}}$ for the GGE $Z^{-1}\Tr (\rho \Or) = \matrixel{\rho_{\rm p}}{\Or}{\rho_{\rm p}}$. The quasiparticle density function $\rho_{\rm p}(\theta)$ is fixed by the thermodynamic Bethe ansatz (TBA) \cite{Yang-Yang-1969,TakahashiTBAbook,ZAMOLODCHIKOV1990695}. In particular, it is related to the state density $\rho_{\rm s}(\theta)$ by the integral equation
\begin{equation}
    2\pi\rho_{\rm s}(\theta)=\partial_\theta p(\theta)+\int d\gamma\,\p_\theta\phi(\theta-\gamma)\rho_{\rm p}(\gamma),
\end{equation}
where $\phi(\gamma) = -i\log S(\gamma)$ is the scattering phase, and $S(\theta_1-\theta_2)$ is the two-body scattering matrix element. Particle and hole excitations on top of the state $\rho$ are denoted $\ket{\rho_{\rm p};\vec{p},\vec{h}}$, where $\vec{p}$ (resp.~$\vec{h}$) are sets of particle (resp.~hole) rapidities, and the associated form factors in interacting integrable models have been studied \cite{denardisParticle2018,CortesPanfil2019,CortesPanfil2019Gen}. These states diagonalize the operators $\vec{Q}=(P,H,\{Q_{i\ge 3}\})$ with one-particle eigenvalues $\vec{\eta}=(\kappa(\theta),\varep(\theta),\{\eta_{i\ge 3}(\theta)\})$. There is a convenient expression for these eigenvalues in terms of the backflow function $F(\gamma|\theta)$ defined by
\begin{equation}
    2\pi F(\gamma|\theta)=\phi(\gamma-\theta)+\int d\gamma'\,\p_\gamma\phi(\gamma-\gamma')\frac{\rho_{\rm p}(\gamma')}{\rho_{\rm s}(\gamma')}F(\gamma'|\theta),
\end{equation}
The eigenvalues of a single excitation are related to the free particle eigenvalues $\vec{h}=(p(\theta),E(\theta),\{h_{i\ge 3}(\theta)\})$ by the relation (see e.g.~\cite{dNBD2})
\begin{equation} \label{dressing}
    \vec{\eta}(\theta)=\vec{h}(\theta)-\int d\gamma\,F(\gamma|\theta)\frac{\rho_{\rm p}(\gamma)}{\rho_{\rm s}(\gamma)}\partial_\gamma \vec{h}(\gamma).
\end{equation}
In addition, the eigenvalues of a state with many particle excitations $\vec{p}$ and hole excitations $\vec{h}$ are additive
\begin{equation}
    \vec{\eta}(\vec{p},\vec{h})=\vec{\eta}(\vec{p})-\vec{\eta}(\vec{h})=\sum_{\theta\in\vec{p}}\vec{\eta}(\theta)-\sum_{\theta\in\vec{h}}\vec{\eta}(\theta).
\end{equation}
The spectral expansion of \eqref{spec1} can now be written explicitly in terms of particle and hole excitations
\begin{equation} \label{spec2}
    \partial_{\lambda^2t}\exv{q_i(0,t)}=2\int d\tilde{\vec{p}}d\tilde{\vec{h}}\,\eta_i(\vec{p},\vec{h})\delta(\kappa(\vec{p},\vec{h}))\frac{\sin\left(\varep(\vec{p},\vec{h})t\right)}{\varep(\vec{p},\vec{h})}|\matrixel{\rho_p}{v}{\rho_p;\vec{p},\vec{h}}|^2,
\end{equation}
with the integration measure given by
\begin{equation}
    d\tilde{\vec{p}}d\tilde{\vec{h}}=\sum_{N=0}^\infty\frac{1}{N!}\sum_{N_p=0}^N\left[\prod_{i=1}^{N_p}\rho_{\rm h}(p_i)dp_i\right]\left[\prod_{j=1}^{N-N_p}\rho_{\rm p}(h_j)dh_j\right],
\end{equation}
where we have split the sum over all $N$ particles involved into the number of particles $N_p$ and the number of holes $N_h=N-N_p$. In addition we have defined the hole density $\rho_h=\rho_s-\rho_p$.

In the Boltzmann regime, obtained by taking the limit $t\to\infty$, the GGE is re-established at all times. We may use $2\sin(\varep t)/\varep \to 2\pi \delta(\varep)$, as well as  the relation $\exv{q_i}=\int d\theta\,\rho_p(\theta)h_i(\theta)$ to take the functional derivative of equation \eqref{spec2} and find the kinetic equation for the distribution function, as given by Eq.~(4) in the main text, with the measure
\begin{equation}
    d\vec{p}d\vec{h}=\sum_{N=0}^\infty\frac{1}{N!}\sum_{N_p=0}^N\prod_{i=1}^{N_p}dp_i\prod_{j=1}^{N-N_p}dh_j.
\end{equation}
Finally,  we can also expand the matrix $\mathcal B$ in Eq.~(10) of the main text in form factors, giving
\begin{align}
    \mathcal B_{ij}=&\int d\tilde{\vec{p}}d\tilde{\vec{h}}\,\eta_i(\vec{p},\vec{h})\eta_j(\vec{p},\vec{h})\delta(\kappa(\vec{p},\vec{h}))\delta(\varep(\vec{p},\vec{h}))|\matrixel{\rho_p}{v}{\rho_p;\vec{p},\vec{h}}|^2,
\end{align}
which can be used to determine the exponential decay of correlations in the final state by Eq.~(12) of the main text.
\section{An H-Theorem}
We now show that the general formalism, independent of any spectral expansion, predicts entropy maximization with respect to the conserved quantities at long times. For a function which decays sufficiently rapidly at infinity we have
\begin{equation}
    \int_{-\infty}^\infty \dd s\,f(s)=\lim_{t\rightarrow \infty}\frac{1}{2t}\int_{-t}^t\dd s\dd s'\,f(s-s').
\end{equation}
Inserting this twice into Eq.~(3) in the main text, for the time and space integrals, gives the following expression
\begin{align}
    &\partial_{\b t}\expectationvalue{q_i}_{\bm{\beta}(\b t)}=\lim_{T\rightarrow \infty}\frac{1}{2T}\int_{-T}^T \dd s \dd s'\lim_{L\rightarrow \infty}\frac{1}{2L}\int_{-L}^L \dd x \dd x'\nonumber \\&\times\bigg[\expectationvalue{v(x,s)Q_iv(x',s')}_{\bm{\beta}(\b t)}-\expectationvalue{Q_iv(x,s)v(x',s')}_{\bm{\beta}(\b t)}\bigg].
\end{align}
Here, the time-evolutions are with respect to the unperturbed Hamiltonian $H_0$. We consider fixed $T$ and $L$, and denote $W^{T,L} = \frc1{\sqrt{4T L}}\int_{-T}^T\dd t\, \int_{-L}^L \dd x\,v(x,t)$. In the GGE state with $\rho(\b t)=Z^{-1}\exp(-\sum_i\beta_iQ_i)$, we can expand in a complete basis of common eigenstates of the charges to obtain
\begin{equation} \label{q_i_intermediate}
    \partial_{\b t}\exv{q_i}_{\bm{\beta}(\b t)}=\lim_{T,L\rightarrow \infty}\sum_{m,n}\left(\rho_m(\b t)-\rho_n(\b t)\right)\left(Q_{i,n}-Q_{i,m}\right)|W^{T,L}_{mn}|^2.
\end{equation}
The von-Neumann entropy density $s = (2L)^{-1}\exv{\log\rho}$ has the time-derivative
\beq\label{tims}
	\p_{\b t} s({\b t})=\sum_i\beta_i({\b t})\partial_{\b t}\expectationvalue{q_i}_{\bm \beta({\b t})}.
\eeq
Using the relation $\sum_i\beta_iQ_{i,m}=-\log \rho_m-\log Z$ and the symmetry $|W_{mn}^{T,L}| = |W_{nm}^{T,L}|$, we obtain
\begin{equation} \label{H-theorem}
   \p_{\b t} s({\b t})=\lim_{T,L\rightarrow \infty}\sum_{m,n}(\rho_m-\rho_n)(\log \rho_m-\log \rho_n)|W^{T,L}_{mn}|^2\ge 0,
\end{equation}
by monotonicity of the $\log$ function.

Equation (10) in the main text also follows from the intermediate result \eqref{q_i_intermediate} if we consider $\rho$ near the stationary point $\rho^0\propto\exp(- \beta_{\text{stat}} (H_0-\nu_{\text{stat}}P))$. Stationarity requires that $\rho^0_m=\rho^0_n$ for any $m,n$ where $\lim_{T,L\to\infty}|W_{mn}^{T,L}|^2$ is non-vanishing. Near the stationary state we therefore have
\begin{equation}
    \partial_{\b t}\exv{q_i}_{\bm{\beta}(\b t)}=\lim_{T,L\rightarrow \infty}\sum_{m,n}\sum_j\beta_j\rho^0_m\left(Q_{j,n}-Q_{j,m}\right)\left(Q_{i,n}-Q_{i,m}\right)|W^{T,L}_{mn}|^2,
\end{equation}
where the linearised density matrix $\rho\propto\rho^0\left(1-\sum_j\beta_jQ_j\right)$ due to commutativity of the charges. One can show that this equation of motion is equivalent to
\begin{equation}
    \p_{\b t}\exv{q_i}_{\bm{\beta}(\b t)}=\sum_j\left([v,Q_i],[v,Q_j]\right)\beta_j,
\end{equation}
with the inner product on the right hand side evaluated in the stationary state. Using the definition of $\exv{q_i}_{\vec{\beta}(\b t)}$ and the chain rule we find that $\dot{q}_i=-\sum_jC_{ij}\dot{\beta}_j$, with $C_{ij}=\exv{q_iQ_j}^{\rm c}$ the static charge correlation matrix, reproducing Eq.~(10) in the main text. Using \eqref{tims} we also recover the first equality of Eq.~(11) in the main text
\beq \label{eqn_entropy}
	\p_{\b t} s = \sum_{i,j\ge 3} \beta_i([v,Q_i],[v,Q_j])\beta_j,
\eeq
where we use the fact that $[v,Q_1]=[v,Q_2]=[h_0,Q_i]=0$ on $\mathcal H''$. Therefore, $([h_0 + \lambda v,Q_i],[h_0 + \lambda v,Q_j]) = \lambda^2 ([v,Q_i],[v,Q_j])$. Writing the operator $\log\rho = -\sum_i \beta_i Q_i - \log Z$, we thus have that the time derivative of the entropy has the suggestive expression, in the Boltzmann regime,
\beq
	\p_t s = ([h,\log\rho],[h,\log\rho]) = ||[h,\log \rho]||^2,
\eeq
where $h=h_0 + \lambda v$ is the perturbed Hamiltonian density, in agreement with Eq.~(11) in the main text.
\section{The Thermal State from TBA is Stationary}
In the preceding section we showed that the long-time stationary state in the Boltzmann regime maximises entropy with respect to the conserved quantities of the perturbed system. We now wish to show directly from the kinetic equation (Eq.~(4) in the main text),
\begin{align}\label{expansionbolt}
	&\partial_{{\b t}}  \rho_{\rm p}(\theta) =
	\int \dd \vec{p} \dd \vec{h}\,K(\theta,\vec{p})B(\vec{p}\rightarrow \vec{h})\big[\rho_{\rm h}(\vec{p})\rho_{\rm p}(\vec{h})-\rho_{\rm p}(\vec{p})\rho_{\rm h}(\vec{h})\big], 
\end{align}
that the stationary state is precisely the thermal state from TBA. Consider the last term in the square brackets, where we write $\rho_{\rm p}=\rho_{\rm s}n$ and $\rho_{\rm h}=\rho_{\rm s}(1-n)$; a similar argument holds for statistics which are not fermionic. A sufficient condition for stationarity is that
\begin{equation} \label{cancel}
    \prod_{p\in\vec{p}}n(p)\prod_{h\in\vec{h}}(1-n(h))-\prod_{h\in\vec{h}}n(h)\prod_{p\in\vec{p}}(1-n(p))=0 \;,\;\forall\,\vec{p},\vec{h}.
\end{equation}
Taking the logarithm, we can ensure this condition is satisfied for all sets $\vec{p}$ and $\vec{h}$ if
\begin{equation}
    \log(\frac{n(\theta)}{1-n(\theta)})=\beta \varep(\theta)+\nu\kappa(\theta),
\end{equation}
as here the energy and momentum delta functions contained in $B(\vec{p}\rightarrow \vec{h})$ (see Eq.~(5) in the main text) ensure the condition \eqref{cancel} is satisfied. We thus find
\begin{equation}
    n(\theta)=\frac{1}{1+e^{\tilde{\ep}(\theta)}},\quad \t\ep(\theta)
    = \beta \varep(\theta)+\nu\kappa(\theta).
\end{equation}
The TBA boosted thermal state is $n(\theta)=(1+e^{\ep(\theta)})^{-1}$, with $\ep(\theta)$ satisfying the integral equation
\begin{equation}\label{ep}
    \ep(\theta)=\beta (E(\theta)+\nu p(\theta))-\int \frac{\dd\gamma}{2\pi}\p_\theta\phi(\theta-\gamma)\log(1+e^{-\ep(\gamma)}).
\end{equation}
Differentiating this equation and integrating by parts we find the following relation:
\begin{align}
    \partial_\theta\ep(\theta)
    =&\beta\partial_\theta  (E(\theta)+\nu p(\theta))+\int \frac{\dd\gamma}{2\pi}\p_\theta\phi(\theta-\gamma)n(\gamma)\partial_\gamma\ep(\gamma).
\end{align}
where $w(\theta) = \beta E(\theta)+\nu p(\theta)$. This is equivalent to relation \eqref{dressing} for $\tilde{\ep}$ \cite{dNBD2}, and thus $\ep(\theta)=\tilde{\ep}(\theta)$ up to a constant. Assuming that this constant vanishes, we have the desired result.

\section{Nearly-Integrable Perturbations}
As discussed in the main text, if we take a current operator $v(x)=j_k(x)$ as the perturbation, then it is nearly integrable in the sense that
\begin{equation}
    \lim_{t\rightarrow \infty}\p_t \exv{q_i(0,t)}=0\;\;\forall \,i.
\end{equation}
To prove this we first use the definition of the projection onto conserved quantities to write
\begin{eqnarray}
    \int \dd x\,\mathbb P[v^0(x,t),Q_i]&=&
    \int \dd x\,\sum_{j,l}\exv{[v^0(x,t),Q_i]q_j}^cC^{-1}_{jl}Q_l \nonumber\\
    &=&\int \dd x\,\sum_{j,l}\exv{v^0(x,t)[Q_i,q_j]}^cC^{-1}_{jl}Q_l
    \nonumber\\
    &=&\sum_{j,l}\exv{v^0(0,t)[Q_i,Q_j]}^cC^{-1}_{jl}Q_l=0,
    \label{projectaway}
\end{eqnarray} 
where $C_{jl}=\exv{q_jQ_l}^{\rm c}$ is the static charge correlation matrix. This immediately shows the equality between Eqs (3) and (8) in the main text. Using the fact that the GGE state should satisfy the cluster decomposition principle at large distances, and using the continuity equation, we also have
\begin{equation}
    V^0(t)=-\int dx\,x\p_xj_k^0(x,t)=\int dx\,x\p_tq_k^0(x,t)=:\p_tE_k^0(t).
\end{equation}
By the results of Ref.~\cite{DY} we have that $[E_k^0(t),Q_i]=\int dx\,\Or_{k,i}^0(x,t)$ for some local operator $\Or_{k,i}$. Hence
\beqa
	( [v,Q_i], \Or) &=& \lim_{T\to\infty}
	\int_{-T}^T \dd t\,\int \dd x\,\expectationvalue{ [j_k^0(x,t),Q_i]\Or(0)}^{\rm c} \n
	&=& \lim_{T\to\infty}
	\int_{-T}^T \dd t\,\expectationvalue{ [V^0(t),Q_i]\Or(0)}^{\rm c} \n
	&=& \lim_{T\to\infty}
	\expectationvalue{ [E_k^0(T)-E_k^0(-T),Q_i]\Or(0)}^{\rm c} \n
	&=& \lim_{T\to\infty}
	\int \dd x\,\expectationvalue{ \big(\Or_{k,i}^0(x,T)-\Or_{k,i}^0(x,-T)\big)\Or(0)}^{\rm c}\n
	&=&0,
\eeqa
for every local operator $\Or$. In the last step we used the hydrodynamic projection principle \cite{DoyonDiffusion2019}, which implies that the large positive and negative time limits are equal. Setting $\Or=j_k$ we obtain $\p_{\bar{t}}\exv{q_i}=0$ $\forall \,i$.

\hspace{-14pt}As discussed in the main text, there exists an orthogonal decomposition
\beq
	\mathcal H'' = \mathcal H_{\rm N} \oplus \mathcal H_{\rm B},
\eeq
into a nearly-integrable component $\mathcal H_{\rm N}$ and a thermalizing component $\mathcal H_{\rm B}$. This is obtained as follows. The subspace $\mathcal H_{\rm N}\subset \mathcal H''$ is defined by the relation written in the main text:
\beq
	b \in \mathcal H_{\rm N} \quad\mbox{iff}\quad ([b,Q_i],a)=0 \quad \forall \quad a\in\mathcal H'', \ \forall i.
\eeq
By invariance of the state with respect to the action of $Q_i$ (that is, the fact that the density matrix commutes with $Q_i$), this implies
\beq
	b \in \mathcal H_{\rm N} \quad\mbox{iff} \quad (b,[Q_i,a])=0 \quad \forall \quad a\in\mathcal H'', \ \forall i.
\eeq
We define $\mathcal H_{\rm B} = \{[Q_i,a]:a\in\mathcal H'',i\}$ as the union of the images of all actions $[Q_i,\cdot]$ of the $Q_i$'s on $\mathcal H''$. We can then see that $\mathcal H_{\rm B}$ is indeed orthogonal to $\mathcal H_{\rm N}$. To make this argument mathematically rigorous requires consideration of the completions of these spaces. This can be done by considering the one-parameter unitary groups generated by the actions of the $Q_i$'s. Note finally that the subspace $\mathcal H_{\rm N}$ is referred to as the ``diffusive subspace", and denoted $\mathcal H_{\rm dif}$, in \cite{DoyonDiffusion2019}. 

In summary, the operators that are responsible for the diffusive effects
of integrable models are also those that, seen as perturbations, do not
lead to thermalization at weak coupling. The rest, corresponding to the
orthogonal complement, are perturbations leading to thermalization. This
is the Hilbert space expression of the separation between diffusive and
thermalizing processes.